*Article*

# Analyzing the Adoption Challenges of the Internet of Things (IoT) and Artificial Intelligence (AI) for Smart Cities in China

Ke Wang [1,*], Yafei Zhao [2], Rajan Kumar Gangadhari [3] and Zhixing Li [4]

1. Department of Civil and Architectural Engineering, Qingdao University of Technology, Shandong 273400, China
2. Building Information Technology Innovation Laboratory (BITI Lab), Solearth Architecture Research Center, Hong Kong 999077, China; yzh@solearth.com
3. Industrial Engineering and Manufacturing Systems, National Institute of Industrial Engineering, Mumbai 400087, India; Rajan.Gangadhari.2018@nitie.ac.in
4. School of Design and Architecture, Zhejiang University of Technology, Zhejiang 310023, China; zxlee910@zjut.edu.cn
* Correspondence: wangke@qutl.ac.cn

**Abstract:** Smart cities play a vital role in the growth of a nation. In recent years, several countries have made huge investments in developing smart cities to offer sustainable living. However, there are some challenges to overcome in smart city development, such as traffic and transportation management, energy and water distribution and management, air quality and waste management monitoring, etc. The capabilities of the Internet of Things (IoT) and artificial intelligence (AI) can help to achieve some goals of smart cities, and there are proven examples from some cities like Singapore, Copenhagen, etc. However, the adoption of AI and the IoT in developing countries has some challenges. The analysis of challenges hindering the adoption of AI and the IoT are very limited. This study aims to fill this research gap by analyzing the causal relationships among the challenges in smart city development, and contains several parts that conclude the previous scholars' work, as well as independent research and investigation, such as data collection and analysis based on DEMATEL. In this paper, we have reviewed the literature to extract key challenges for the adoption of AI and the IoT. These helped us to proceed with the investigation and analyze the adoption status. Therefore, using the PRISMA method, 10 challenges were identified from the literature review. Subsequently, determination of the causal inter-relationships among the key challenges based on expert opinions using DEMATEL is performed. This study explored the driving and dependent power of the challenges, and causal relationships between the barriers were established. The results of the study indicated that "lack of infrastructure (C1)", "insufficient funds (C2)", "cybersecurity risks (C3)", and "lack of trust in AI, IoT" are the causal factors that are slowing down the adoption of AI and IoT in smart city development. The inter-relationships between the various challenges are presented using a network relationship map, cause–effect diagram. The study's findings can help regulatory bodies, policymakers, and researchers to make better decisions to overcome the challenges for developing sustainable smart cities.

## 1. Introduction

The rapid migration of people to urban areas from rural areas is causing a burden on big cities. The city councils and local governments are facing issues in managing the essential needs of huge populations. On the other hand, global warming, climate change and technology's move towards renewable energy has pushed the revolution of smart cities, which aim to provide all essential needs to the public without causing many negative impacts on the environment. With the rapid development of science and technology today, technological innovation has become a leading force in driving economic and social development. These advanced cities must use new technologies to improve their core systems to maximize optimization in all functions of city management by using limited energy. As an important carrier of economic and social development, cities are also the main gathering place of innovation factors, and the role of science and technology innovation in urban development is becoming increasingly prominent, and is becoming the engine of future development of cities [1]. Due to the recent industry achievements, the Internet of Things (IoT) and artificial intelligence (AI) are current popular research topics, and they have proven to return better results in many disciplines such as automating factories, public surveillance, asset monitoring, waste management, weather monitoring, etc. Combining IoT and AI is an effective way to intelligently upgrade the existing information systems.

Global economic crises can often lead to technological revolutions. For example, the world economic crisis in 1857 led to the first technological revolution, the world economic crisis in 1929 led to the second technological revolution, and the economic crisis in 1987 led to the information technology revolution [1]. After the financial crisis in 2008, smart cities have become a global trending topic (Figure 1) with the iconic report Smart Planet by IBM. Smart Planet advocates the full use of next-generation information technology in all industries [2], thus giving rise to concepts such as smart cities and digital cities. In the present day, the concept of smart cities is not clearly and statically defined [3]. However, it is clear through its interdisciplinary development that smart cities are deeply integrated with information and communication technology (ICT) and IoT [4]. The objective of smart cities is to enhance the efficiency of resource utilization, optimize city management and services, and improve the quality of life of citizens [5]. A smart approach with the help of AI and the IoT can be applied to smart transportation, security, energy, buildings, education, health, and more [4,5]. The smart city framework contains a large number of indicators to measure sustainability, such as social and economic sustainability, and AI and IoT technologies also help to maintain sustainability, so Khan et al. focused on the sustainable smart city and analyzed the challenges it faces [6]. Treude summarized the definitions proposed by several scholars for the sustainable smart city. He argued that finding a comprehensive, sustainability-oriented definition of a smart city is a complex challenge. Treude, therefore, proposed a limited definition: a city is smart if it uses smart technologies to better address 21st-century challenges. If it includes sustainable development goals, then it represents sustainable urban development [7].

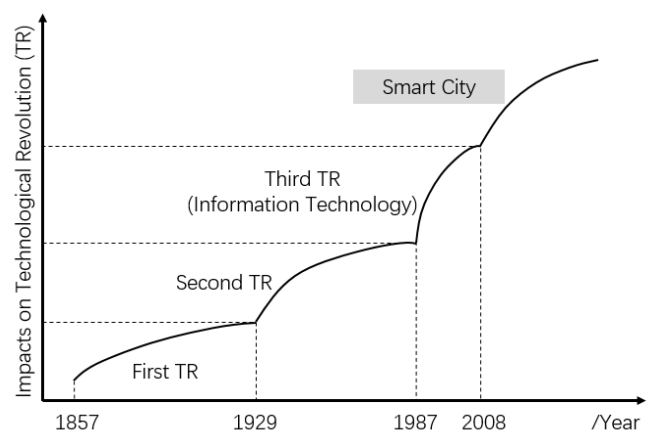

**Figure 1.** Impacts on technological revolutions by global economic crises (source: IBM, 2008).

Cities continue to attract new populations, and the United Nations estimates that by 2030, more than 60% of the global population is expected to live in large cities [8,9]. Population migration to large cities has become a common phenomenon worldwide. For example, the current resident populations of Beijing, Shanghai, and Shenzhen are all more than 20 million [10]. Huge population growth brings many impacts and challenges to urban resources and services [8]. An analysis of public reports and government documents showed that in recent years, there has been an increasing number of smart city pilot projects in China (Figures 2 and 3). The increasing trend is indicated in Figure 2, which shows the number of China's smart city pilot projects, and summarizes the data from different years and integrates them into one form. The data for different years were from several sources, which are listed below the title of figure. They are China MOHURD, China NDRC, and Forward the Economist. China MOHURD is the Ministry of Housing and Urban–Rural Development of China. China NDRC is National Development and Reform Commission. Forward the Economist is one of the popular media and academic report analysis publishers in China. Sources provided by MOHURD and NDRC are government documents, while sources by forwarding the Economist are public reports. Refs. [11–18] are the lists of original data published by the cited source that are integrated into the figure. The original data from the above sources are separate, therefore the figure summarizes them and verifies the increasing trend.

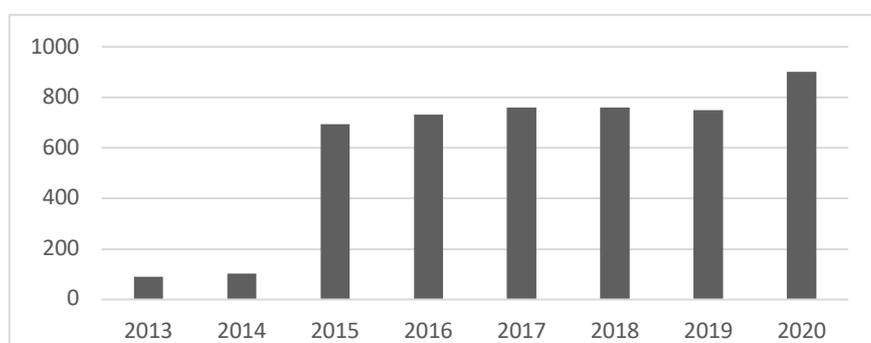

**Figure 2.** Number of smart cities with experimental policies in China (data source: China MOHURD, 2013, 2015; China NDRC, 2020; Forward the Economist, 2019, 2020, 2021).

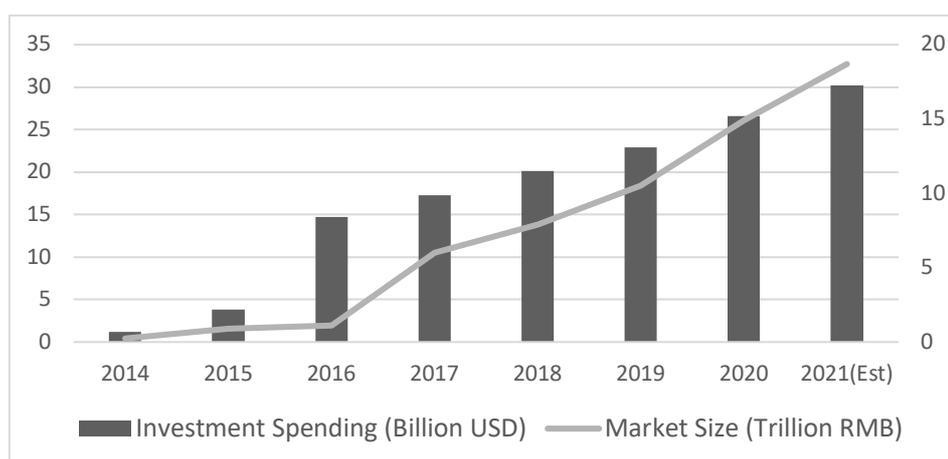

**Figure 3.** Trends for market size and investment spending of smart cities in China (data source: CCIT, IDC Qianzhan Industry Research Institute, 2019, 2020, 2021).

The idea of "smart" is to use information technology to drive the operation of the city, which includes monitoring, forecasting, and real-time management. The combination of IoT and AI can replace the traditional means of managers in the past, with IoT mainly referring to sensors or hardware, and AI mainly referring to back-end algorithms. With the advancement of technology and the increasing awareness of smart city residents, the concept of smart is no longer limited to intelligence, but fully incorporates the participation of residents or users. The mode of engagement includes not only citizen participation, but also decision making. Decisions must be made in a fast and effective manner, often relying on real-time feedback and behavioral changes from citizens [19]. Decisions are made by top-down or bottom-up approaches, are technology or community-led, and include purposeful design or natural evolution. In well-functioning cities, citizens' decisions are dependent on multiple types of tools that guide them in their daily decision making [20]. According to IDC's *Worldwide Smart City Spending Guide* released in July 2020, China's smart city market spending will reach USD 25.9 billion in 2020, up 12.7% year-on-year, higher than the global average, and the second-largest spending country after the United States [21].

In recent years, due to the improvement of cloud computing technology, computers have been able to overcome the problems that used to occur due to the lack of arithmetic power. Artificial intelligence algorithms are being noticed by several industries due to the development of cloud computing, including the study of smart cities. Software, applications, and plugins are usually deployed on a cloud server and operated with cloud computing techniques. According to the investigation, for the Infrastructure as a Service (IaaS) category, the top five companies in the Chinese public cloud computing market are Alibaba Group (Aliyun), Tencent, China Telecom, Kingsoft, and Amazon Web Services (AWS). However, for the Platform as a Service (PaaS) category, the top five companies in the Chinese public cloud computing market are Alibaba Group (Aliyun), Oracle, Amazon Web Services (AWS), Microsoft, and IBM [22]. Wu summarized that from a functional point of view, the smart city system can be divided into a perception layer, a network layer, and an application layer [1]. According to Cohen, there are three stages in the development of smart cities: 1.0 Technology-Driven, 2.0 Technology-Enabled, and 3.0 Citizen Co-Creation. Different countries and cities are currently at different stages [23], and therefore they have different levels of adoption. The level of adoption reflects the degree of acceptance of technology in an object. It is a series of driving processes, from awareness to acceptance to use. This study aimed to examine the challenges hindering the adoption of AI and the IoT in smart city development by evaluating the current literature using the PRISMA method and incorporating expert opinions to select the appropriate barriers. Using the DEMATEL model, we developed cause-and-effect relationships between key AI and IoT implementation challenges. The findings of the study will be useful to understand the cause–effect relationships between the challenges that would help policy makers, practitioners, and researchers to understand the effect of the challenges in building smart cities in China.

## 2. Development of AI and IoT for Smart Cities

### 2.1. The Case for Smart City Development

Singapore is considered a pioneer in the smart city movement (Qi and Shen, 2019). Emerging technologies of IoT are used for the development of smart cities in smart fleet management, air quality monitoring, energy management, and smart agriculture needs with the help of sensors, robots, and other cyber–physical systems. All the generated data in this process are sent to the cloud (a central platform) for processing, and the output generated is used to plan smart city strategies [24].

In the context of Industry 4.0, the combination of smart cities with IoT and AI can lead to win–win results. Tobias' research highlighted the importance of monitoring systems [25]. In New York, for example, the government has upgraded its lighting system to enable intelligent control of dimming and turn-on times. It is also reducing lighting power by retrofitting LED installations. An automated meter reading (AMR) system is used to provide feedback on water usage, and also to monitor potential leaks in real time. Tobias also described the representative smart waste management, a technology that is also widely used in Singapore. As part of the smart waste management program, monitors on the bin lids collect information about the contents and location, and transmit the information to the waste team via a central server. The route of the waste collection team can then be optimized [26]. In Singapore, the Smart Nation vision was established in 2014 to harness new technologies to address growing urban challenges. To date, the most developed smart services in Singapore are the intelligent transport system (ITS), which is more than 10 years old, and e-government, which has been developed since the early 1980s [5]. In addition, One Monitoring, developed by the transport department, provides services to all drivers and vehicle owners in the city, including access to information about cabs on the road and real-time road information, among other features [26].

In China, the Premier's report "Making Science and Technology Lead China's Sustainable Development" in 2009 opened the era of smart cities. Over the past decade, the smart city industry chain has involved interdisciplinary and multi-industries, integrating multilevel technologies. Based on the upstream and downstream industry chains of smart cities described by [21,27–29], this study integrated the industry ecosystems and technology architectures involved in smart cities. Figure 4 illustrates the streams (up, middle and down) for the integrated structure of industrial chains of a smart city. The vertical axis shows multiple layers, including upper terminal device, platform service, and the bottom infrastructure.

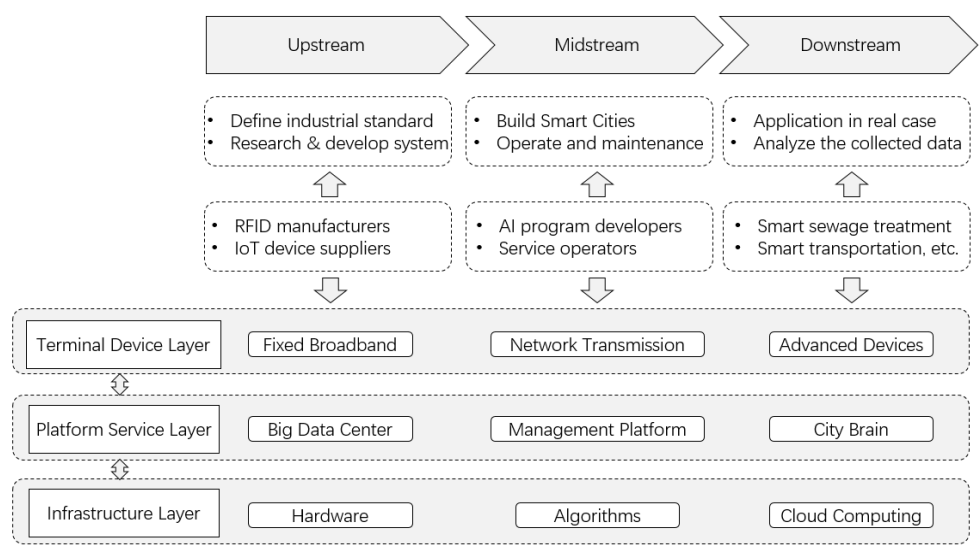

**Figure 4.** Integrated structure of industrial chains of a smart city (data source: Qianzhan Industry Research Institute, 2019, 2020, 2021).

As of 2019, there were more than 700 smart city pilots in China. The market size is growing by more than 30% per year, reaching RMB 10.5 trillion in 2019. Of these, up to 94% of provincial-level cities and 71% of prefecture-level cities have launched smart city programs [30].

Hangzhou proposed the City Brain system in April 2016. It is a digital interface for citizens, including 11 systems and 48 application scenarios such as police, transportation, cultural tourism, and health, with an average of more than 80 million data points per day [31]. The City Brain connects and shares data resources that were originally scattered in various departments and isolated from each other. By establishing a high-speed urban "CPU", the government's service effectiveness is continuously improved [31]. The City Brain has taken over the signals of 1300 intersections in a 420 km$^2$ area of Hangzhou. The functions that have been realized to facilitate people's livelihood include convenient parking, digital tourism, smart environmental protection, vegetable-planting base management, convenient access to medical care, and many other application scenarios. Therefore, the use of cutting-edge technologies such as big data, cloud computing, blockchain, and artificial intelligence to promote urban management means and management modes is a way to drive cities from digitalization to intelligence [30].

Shenzhen proposed the concept of "Building a Smart Shenzhen" in 2010, and is now one of the representative smart city models in China. Shenzhen is a mega city with more than 20 million people, so the scale of data and the way it is processed is very important. In the dimension of a smart city, "smart" means to have data, and that the data is connected [32]. To improve the efficiency of urban governance and industrial sectors, it allows the sharing of effective data resources across sectors, industries, and fields, so that these integrated data and new technologies can improve capacity, output, and efficiency. For example, the Construction Digital Management Center (CDMC) was developed for metro construction sites. It enables centralized control of more than 400 construction sites across the city and uses AI algorithms to identify safety hazards, greatly enhancing work efficiency [32].

*2.2. Features of the IoT*

The Internet of Things (IoT) is a highly integrated and comprehensive use of next-generation information technology, which is important for a new round of industrial transformation and green, smart, and sustainable economic and social development [27]. New IoT applications are driving smart city initiatives around the world. The IoT is based on the installation of sensors (RFID, IR, GPS, laser scanners, etc.) for everything and connecting them to the Internet for information exchange and communication through specific protocols for intelligent identification, location, tracking, monitoring, and management [6]. The IoT and IT are at the core of the hyperconnected society, which is also known as Machine to Machine (M2M) or Internet of Everything (IoE) [33]. Theodoridis developed a city-scale testbed for IoT and future Internet experiments [34]. Latre et al. [35] developed the City of Things smart city testbed located in Antwerp, Belgium. The platform consists of a multi-wireless technology network infrastructure and includes an integrated approach that allows experimentation in three different layers: network, data, and living labs [35]. Sanchez deployed the Smart Santander project in the city of Santander, a typical application of the Internet of Things in a smart city [36].

Big data highlight openness and connectivity [37], and the combined application with IoT can moreover stimulate the value behind it. Chen et al. proposed a cloud platform architecture for building energy management based on big data and distributed technology using machine learning and environmental real-time analysis to provide technical support for building operations [38]. Zhang et al. summarized common data mining techniques, such as association mining, multivariable linear regression modeling, classification, clustering, and neural networks, and analyzed the data-mining techniques and the methods and trends of applying various data-mining techniques in the field of building energy efficiency [39]. Shi et al. used the SSIS algorithm with SQL to store historical data on energy consumption, and used Visual Studio as a development tool based on the ASP.NET MVC framework to establish an energy consumption management platform and energy consumption management software [40]. Yang et al. proposed a data-mining-based method for processing energy consumption data from public buildings, including building data sets based on historical data and classifying them according to different energy use patterns [41].

*2.3. Implementation of AI in Smart Cities*

As an important part of the smart city process, the construction industry plays a pivotal role. Technologies based on IoT and artificial intelligence algorithms have greatly accelerated the development of the construction industry. Artificial intelligence algorithms are a means of bulk problem solving [42]. At the same time, intelligent energy simulation tools combining parametric methods, BIM technologies, or computer programming techniques have emerged based on the architectural design perspective [43].

A smart city refers to a regional scale that aggregates multiple buildings. Therefore, the shared data of each building form the interconnection system. Building performance simulation technology can be used to rationalize or reduce the energy consumption of buildings. This method is called sensitivity analysis or parametric simulation. The combination of optimization algorithms to analyze changes in building energy consumption by changing several parameters simultaneously is called building performance optimization [44]. There are two main methods used to predict energy consumption in buildings, the data-driven approach and the physical modelling approach [45]. Due to the development of cloud computing and algorithms, the data-driven approach has received increasing attention in recent years. The common machine learning algorithms are support vector machines (SVMs), artificial neural networks (ANNs), and decision trees.

Amasyali et al. reviewed the research on data-driven building energy prediction models, with a particular focus on reviewing the prediction scope, data attributes, and pre-processing methods in conjunction with machine-learning algorithms. Their study proposed future research directions in the field of data-driven building energy prediction. A brief overview of existing review studies on data-driven building energy prediction was presented by considering building type, time granularity, predicted energy consumption type, data type, element type, and data size [45]. For the machine-learning algorithms used to train energy consumption prediction models, the mainstream ANNs, SVMs, and DTs occupied 76% of the distribution, while 24% of the studies used other statistical algorithms, and some scholars also conducted comparative studies between two or more algorithms to derive similarities and differences in energy consumption prediction (Table 1). Nearly half (47%) of the studies focused on the prediction of overall building energy consumption [45]. A combination of multiple algorithms can be used to efficiently solve the problem of building energy consumption in smart cities. The below table summarizes the literature on building performance simulation and prediction in smart cities based on mainstream algorithms.

**Table 1.** Summary of algorithms mentioned by previous researchers for improving building performance management in smart city projects.

| Scholars | Research Topics | Algorithms or Models |
|---|---|---|
| Sadeghi et al. [46] | Predicting residential building energy performance | DNN |
| Dong et al. [47] | Predicting residential electricity energy consumption | Combination of ANN, SVR, LS-SVM, GPR, GMM |
| Li et al. [48] | Predicting building electricity consumption | Optimized-ANN, PCA |
| Ahmad et al. [49] | Forecasting building electric energy consumption | ANN, SVM |
| Daut et al. [50] | Analysis of building electrical energy consumption prediction | Mixture of SVM and SI |
| Turhan et al. [51] | Comparative study of building heat load estimation | Combination of KEP-IYTE-ESS and ANN |
| Qiu et al. [52] | Simulation of vacuum photovoltaic glass and building energy consumption | Coupling ANN with light harvesting model |
| Zhong et al. [53] | Prediction of energy consumption in office buildings | SVM |
| Roldán-Blay et al. [54] | Prediction of building electric energy consumption | ANN and temperature profile model |
| Goyal et al. [55] | HVAC zone control for occupancy | MPC |
| Williams et al. [56] | Predicting monthly residential energy consumption | MARS |
| Massana et al. [57] | Short-term energy consumption forecasting for nonresidential buildings | MLR, MLP, SVR |
| Shi et al. [58] | Forecasting energy consumption in office buildings | ESN |
| Jovanović et al. [59] | Forecasting heating energy consumption for university campuses | FFNN, RBFN, ANFIS |
| Bourhnane et al. [60] | Smart building energy consumption forecasting and planning | GA, ANN |

**Abbreviations:** Adaptive Neuro-Fuzzy Inference System (ANFIS), Artificial Neural Network (ANN), Deep Neural Network (DNN), Echo State Network (ESN), Feed Forward Neural Network (FFNN), Multivariate Adaptive Regression Splines (MARS), Multilayer Perceptron (MLP), Multiple Linear Regression (MLR), Model Predictive Control (MPC), Principal Components Analysis (PCA), Radial Basis Function (RBF), Swarm Intelligence (SI), Support Vector Machine (SVM), Support Vector Regression (SVR).

## 3. Literature Review and Challenges to IoT and AI Adoption in Smart Cities

The arrival of new technologies is always accompanied by opportunities and challenges. Although the concept of smart cities was only created less than 20 years ago, the rapid iterations and upgrades have brought about many challenges as well. This section identifies and summarizes influencing factors from a systematic literature review of published papers and studies. Preferred Reporting Items for Systematic Reviews and Meta-Analyses (PRISMA) was applied in the review process. PRISMA is a commonly used method that clearly presents the process and reasons of document identification, screening, inclusion, or exclusion, which can improve the accuracy of reviews and meta-analysis reports [61]. The initial source data was collected from major English and Chinese academic search engines, including Web of Science, Engineering Village, Scopus, Google Scholar, CNKI, and Wanfang Data. By using the identified keywords "IoT/ Internet of Things", "AI/ Artificial Intelligence", and "Smart City/ Smart Cities"; and "adoption", "challenge", "barrier" and their relative Chinese translations with Boolean structured searching methods, a total of 439 initial papers were collected. Then, 128 duplicated articles were removed. After reading the title, abstract, and keywords of those collected papers and determining whether they were aligned with the research topic, 87 of them that did not meet the criteria were removed. Then, a second measurement was done to check if the main content of those papers met the criteria, and whether they were articles, reviews, degree theses, or conference proceedings. After this step, 171 articles were included for further categorizing (see Figure 5). The final 171 articles selected based on the above bibliographic research were studied carefully to identify the potential challenges of AI and IoT adoption for smart city development. After analyzing the content of these articles, 10 influencing factors (challenges) were identified, and the supporting literature is presented in the Table 2. The roles and interrelationships of the identified challenges were established using DEMATEL by considering the experts' opinions.

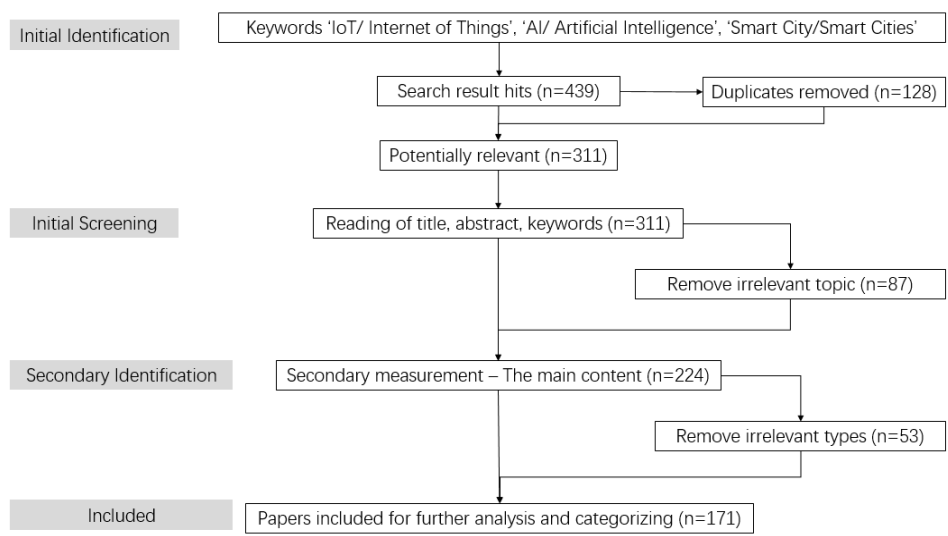

**Figure 5.** Literature search results using the Preferred Reporting Items for Systematic Reviews and Meta Analyzes (PRISMA) method

**Table 2.** Ten key challenges identified in the literature, with sources.

| No. | Challenges (Factors) | Code | Reference |
|---|---|---|---|
| 1 | Lack of Infrastructure | C1 | [62-73] |
| 2 | Insufficient Funds or Capital | C2 | [74-91] |
| 3 | Cybersecurity and Data Risks | C3 | [92-119] |
| 4 | Smart Waste and Hygiene Management | C4 | [120-140] |
| 5 | Lack of Professionals | C5 | [141-156] |
| 6 | Managing Energy Demands | C6 | [157-176] |
| 7 | Managing Transportation | C7 | [177-206] |
| 8 | Environmental Risks | C8 | [207-232] |
| 9 | Managing Public Health and Education | C9 | [233-244] |
| 10 | Lack of Trust in AI and IoT | C10 | [229, 245-265] |

*3.1. Lack of Infrastructure*

Smart cities need the latest very advanced infrastructure, and every piece of equipment should be connected to the Internet to monitor it. In smart cities, connected IoT devices collect data from the physical medium to optimize decisions to improve urban services for citizens [266]. The growth of the population requires the need for housing, educational institutes, hospitals, and entertainment facilities. AI and the IoT should address the needs of infrastructure for all the people living in smart cities that will become part of the IoT infrastructure [267]. Artificial intelligence has important implications for achieving sustainable development, which is directly related to infrastructure development in emerging economies [268]. In addition, AI and the IoT should take care of utilizing and connecting the old (existing) infrastructure for multiple purposes. IoT technologies have contributed significantly to most of the detailed aspects of smart city technologies and infrastructures. As the basic concepts and ideas of IoT technologies are shared with smart city technologies and infrastructures, there are substantial business opportunities and extensive growth potential [269].

*3.2. Insufficient Funds or Capital*

A study showed that funds and budget were the main factors to begin any project [74,270]. The government or local authorities should have sufficient funds to design, develop, and continue smart city planning. Unavailability of funds delays the implementation of projects, which again causes an increase in project costs. Abdalla considers the lack of investment and capital funding as the main threat to smart city strategies [271]. AI and IoT techniques should help in prioritizing the projects based on the severity, requirements, timelines, and other parameters, and optimally allocate the budgets using the best optimization techniques [272-274]. To effectively address this issue, the government may need to obtain additional funding from entities in the private sector that are interested in these smart city projects [275].

*3.3. Cybersecurity and Data Risks*

Smart cities' interconnective networks create huge amounts of data containing rich information; bring innovation; and connect governments, industries and citizens. The data create a foundation for operating cities that will make them more effective and sustainable. Sharing and storing large amounts of actionable data also raises many concerns and challenges. This might include data related to citizens private information, government documents, and the information of all private organizations. On the other hand, cyber insecurity raises concerns about data privacy and threats to smart city systems [273,274]. For example, there was a cyber attack on a top US pipeline network company, and the company was forced to shut down its operations for few days and paid a huge amount to the hackers to resume its operations [274]. On 20 August 2021, China passed a new personal data privacy law that will take effect on 1 November 2021 [275]. This law requires that separate consent be obtained from individuals for the processing of sensitive personal information such as biometric, medical health, financial accounts, and whereabouts. It also prohibits the collection of sensitive information from individuals in certain special cases of writing, such as facial information and biological information. This might create challenges for the tech giants to handle the information without mismanagement and misuse. Due to the increasing volume of sensors and their data, robust connectivity technology is a requirement for success. [245] Without powerful citywide coverage, the success of such a project is more than unlikely. Smart cities will house millions of people, and managing the people data, analyzing it, and preventing cyber-attacks is a challenge for AI and IoT. Therefore, an integrated approach is needed to deal with data security, user privacy, and trust in smart cities [276]. Meanwhile, the challenge of cybersecurity and data risks and the associated responsibilities must be shared by all parties involved in the smart city process, including city managers, residents, and the society itself. Cybersecurity and data risks comprise various types, and a cyber attack is one of the aspects. Therefore, effective measures have to be taken for smart cities to elaborate methods to detect attacks. Pasqualetti et al. have designed a centralized and distributed attack detection and identification monitor by describing monitoring limitations [277]. At the same time, they proposed a framework for cyber–physical systems, attacks, and monitoring. Carl et al. explored the capabilities of DoS attacks by analyzing denial-of-service attack detection techniques, including activity-based analysis, change-point detection, and wavelet-based signal analysis [278]. They argued that the detectors used for testing do not solve the problem independently, and that a combination of methods could produce better results in the future. Salem et al. surveyed insider attack detection and solutions in the literature [279]. Their approach was to classify insider attack detection cases. By summarizing the current common methods and techniques for insider attack detection, future research directions were proposed. Data security is particularly important in the medical field, and Liag et al. developed an attack detection model for medical surveillance systems using Internet Protocol (IP) virtual private networks (VPNs) over optical transport networks. They proposed a multilayer network architecture and a lightweight implementation of the procedure that enabled remote access to medical devices [280].

*3.4. Smart Waste and Hygiene Management*

There is a growing interest in the potential application of these technologies in manufacturing systems and municipal services to ensure flexibility and efficiency [120].

Handling waste still remains a major concern for many countries. Waste management from inception to disposal is one of the key challenges faced by municipal companies worldwide [281]. With the current phase of living styles in which most of the food and other items are wrapped with plastic or paper packaging, handling the waste produced by millions of people is a huge task. Waste collection must be completed within a specified time frame, as cities generate waste at an alarming rate and need to collect it more smartly [282]. Some waste management companies have developed AI-enabled waste segregation techniques that automatically separate different kinds of waste (e.g., paper, plastics, glassware, metals, etc.) automatically without human intervention. AI and the IoT should address the issues associated with collection, transportation, treatment, recycling, and disposal in waste management. For example, the intelligent disposal process was achieved through Ant colony optimization (ACO) technology. The entire process can be monitored centrally, thus providing high-quality services to the citizens of smart cities [283].

*3.5. Lack of Professionals*

The adoption of AI and the IoT require highly skilled professionals. Without proper experience and knowledge, organizations always misunderstand the benefits of these technologies. One of the current challenges facing smart cities is the lack of professionals with knowledge of computer technology and knowledge of different fields [141,142]. The unavailability of skilled professionals lags behind the adoption of AI and the IoT in sustainable smart city development [284]. Novák et al. provided a detailed analysis of the lack of professionals in the Middle East, Amsterdam, and Hungary [284-286]. The attraction and adoption of professionals will be particularly important in the post-pandemic era. The epidemic has already had an economic impact on the city, and if there is a chronic shortage, this will make it difficult to achieve intelligent urban management, and will affect the efficiency and quality assurance of smart city operations. For the construction of smart cities, the knowledge of professionals helps to achieve desired goals [287].

*3.6. Managing Energy Demands*

Smart devices are integrated and converted from everyday objects with advanced computing algorithms, and become the intelligent terminals that transfer data with a cloud server or other devices [288]. Thus, it requires a huge amount of energy to maintain the smart city [289]. Providing energy for a city's needs is a challenging task, and with increasing attention shifting towards renewable energy sources, several cities have struggled in migrating to renewable energy sources, On the other hand, energy demand and costs are increasing over time [290]. Domestic energy consumption is exponentially increasing due to the use of modern televisions, air conditioners, washing machines, smartphones, and computers. The technological advances and changes in consumer habits are leading to higher energy demands [291], and now the energy producers are looking for help from AI and the IoT to optimize the distribution of energy demands through methods such as automating streetlights, increasing unit electricity prices during peak times, and updating old equipment with modern equipment. AI implies new rules for organizing activities. There is a need to improve the design, deployment, and production of energy infrastructure to meet multiple challenges [292]. One solution to this challenge is to install solar power generation facilities for homes to help alleviate the energy consumption problem. Modern homes can also help boost the value of fixed assets by installing auxiliary solar power systems. Since solar panels do not generate electricity at night, homes will begin to draw power from the main grid as usual after the sun goes down. The Mississippi Power Company has announced a partnership with Tesla to create the world's first smart community with a Tesla residential solar+storage system [293]. Anbari argues that the combined use of solar and wind resources in a smart city environment can help energy companies to be able to meet the energy needs of cities [294]. Northfield argues that there is a need to maximize the use of the regional grids powered by solar, wind, or hydro energy so that the growing energy demands can be met locally instead of relying on a central supply line [295].

*3.7. Managing Transportation*

As a noteworthy part of today's economy, transportation accounts for 6–12% of a country's GDP that can be created. Although transportation has greatly improved our lives, many excessive problems remain unresolved [296]. Transportation has become the second-largest carbon-emitting sector due to inefficiencies. Current transportation methods are heavily dependent on crude oil products such as petroleum, diesel, etc. Electric cars are a good alternative to combat the emissions-related problems. Electric vehicles generate the necessary power for driving using a battery pack and electric motors. The charging requirements of electric vehicles can be met if existing fuel stations are upgraded to hybrid modes that provide petroleum products as well as e-charging stations. To avoid traffic congestion, it is important to design a city to minimize the public daily transportation requirement that can bridge the gap to reduce residents' travel times. One of the main branches of smart cities is smart transportation [297]. There is no smart city without a reliable and efficient transportation system. This necessity makes intelligent transportation systems (ITSs) a key component of any smart city concept [298]. This affects not only smart transportation, but also the environment [299]. Saroj et al. developed a real-time, data-driven smart transportation simulation for smart cities. This simulation model was used to evaluate and visualize the feasibility of network performance metrics to provide dynamic operational feedback in a real-world, big data environment [300]. The machine-learning techniques that are an integral part of AI should be capable enough to analyze the past data from public and private transportation activities to analyze the root where frequent congestion occurs or where most accidents occur, the reasons for the accidents, and preventive measures to address these issues. In China, besides the smart taxi dispatch system for automobiles, there are also smart systems for bicycles, which is known as the shared bicycle. To improve the efficiency of shared bicycle use, the algorithm will propose a reasonable scheduling plan based on real-time road conditions, after which it will be carried out manually. Bicycle sharing is a solution to the "last mile" problem for residents, as many users choose to ride bicycles as an alternative to walking short distances.

*3.8. Environmental Risks*

Cities are becoming increasingly vulnerable to environmental risks and climate change [207]. In July 2021, several European countries and China experienced severe floods that caused millions of dollars of property damage and loss of people's lives [207,301]. Smart cities should have highly responsive and agile disaster management systems such as weather monitoring, and alerting people regarding preventive measures to reduce pollution levels. Due to the increase in the number of inhabitants, there is a constant need to support economic growth to the point of creating environmental risks [301]. Large population concentrations also result in an increase in pollution from transportation, single-use plastics, and other types of waste, which largely contributes to environmental pollution [302]. Large-scale housing needs also pose several threats to the environment [305,306]. AI and IoT can be used in the design phase to minimize environmental risks. Automated drones have been used in various fields such as environmental risk detection and capture, traffic regulation, and air pollution monitoring. Air quality sensors on publicly accessible online platforms can support the measurement of environmental risks [305,306]. Urban environments improve competitiveness and respond to environmental risks to make cities smarter [307].

*3.9. Managing Public Health and Education*

By 2020, 86% of health-related companies will spend an average of USD 54 million on artificial intelligence and healthcare [308]. In China especially, which is home to the world's largest population, providing healthcare facilities to all people is a challenge. It is difficult to provide the necessary medical requirements to all age groups without assessing their past health conditions. AI should learn to assess the health condition of a patient using their past medical records and present physical situation. The digitalization of medical records on a central server will help governments to assess the data anytime and anywhere in the country. This would help medical staff to investigate the patients and treat them carefully. Besides using a central server to store the medical records, there are also alternative methods using blockchain. Holbl et al. [309] have studied the applications of blockchain in healthcare, and marked it as an enabler for decentralized healthcare management. Kong proposed an AI-based model for online triage in sustainable smart city hospitals [310]. IoT-enabled medical equipment such as operation theatres, diagnostic tools, and smart equipment should assist in managing health. For example, during the COVID-19 pandemic, countries adopted measures for effective protocols for sharing health data. Allam argued that only by designing different smart city products to support standardized protocols that allow seamless communication between them can better urban structural risk management decisions be provided [246]. The flip side of the sensible use of new technology is the misuse of personal health data. Measures involving data collection and use all inevitably trigger data breaches or misuse. The examples include personal journeys during the COVID-19 pandemic or records of visits for difficult medical conditions, which can compromise the rights of patients or even the general population. A new law enacted in China in August 2021 also prevents such incidents from occurring.

Education is also a primary need for the citizens. AI should help in the design and development of educational programs according to industrial and academic needs. Continuous revision of syllabi, assessment of student skills, and analyzing the industrial job requirements of developing skills in students are some examples in which AI can play a role. In this era of data abundance and exponential growth of new knowledge development, IoT is challenging institutions to rethink teaching and learning in the global marketplace [311]. The use of IoT can exchange and utilize information in very appropriate ways to facilitate student engagement and interaction with their peers and teachers. Mahmood explored the applicability of Raspberry Pi development boards or single-board computers in teaching enhancement, IoT technologies, and environments, and proposed a low-cost, efficient, and flexible educational platform [312]. Nuseir focused on entrepreneurial intentions in smart cities and elucidated the relationship between entrepreneurial competencies, entrepreneurial self-efficacy, and entrepreneurial intentions [247].

*3.10. Lack of Trust in AI and IoT*

Trust is generally located at the level of interpersonal relationships. In modern life, more particularly in smart cities, trust between people is increasingly systemic trust [248]. The lack of trust in AI and the IoT might slow their implementation in smart city development. As mentioned by researchers in [313,314], building the trust AI models that can transform social, political, and business environments and can help people in decision-making processes will remove any negative opinions about the usage of AI and IoT technologies. Government agencies should try to create awareness by spreading positive news about using AI in sustainable practices, maintaining transparency, and publishing some case studies that might improve the trust. The report "Artificial Intelligence and Life in 2030: One Hundred Year Study on Artificial Intelligence" by Stone et.al analyzed in detail how AI could impact a typical North American city in 2030 [315]. Each of these chapters and predictions for the next 15 years reflect the different AI impacts and challenges. Examples include the difficulty of creating secure and reliable hardware, the challenge of gaining public trust, and the social and societal risks of reduced human interaction. Poola mentioned simple tasks that AI can perform, such as facial recognition and car driving, as well as complex tasks such as developing a super AI that improves itself and triggers an intelligence explosion. It is also possible to use advanced technologies to eradicate poverty while also developing appropriate precautions for AI [316]. Kamble et.al provided an overview of AI and its applications in human life [317]. For example, they explored the current use of AI techniques in cyber intrusions to protect computers and communication networks from intruders. In the medical field, technologies are used to improve hospital inpatient care. Smart city data management systems provide the collected data and generate revenue, but the system should also maintain people's trust while doing so [318]. Managing and building people trust is a key challenge in sustainable smart city development. Data- and citizen-centric smart city governance is essentially built trust [319]. However, Falco proposed the concept of participatory AI, and he argued that engaging the public will not only increase community trust in AI [249]. Hwang analyzed the evolution of smart cities in Korea. He argued that many Koreans have lost trust in smart city development due to their experiences with u-City [250]. However, the birth of new technologies can change this situation to some extent. The contribution made by the AEC (Architecture, Engineering, Construction) industry is indispensable to the process of building smart cities. Solearth and Zhao proposed the discipline of AEC-IT that integrates the AEC industry with advanced Information Technology in the era of the AI age [320]. Solearth and Zhao proposed a new view that currently, the stage of the AEC industry or architectural design sector is in the second half of the industry cycle, comparing with hundreds of years' development. AI and IoT play important roles in the process of AEC enterprises, from the original purely design service to DBOB mode (Design, Build, Operate, Brand) [320].

This section uses a systematic literature review approach to discuss the use of AI and IoT in smart cities. A total of 10 challenges were identified in the literature using the PRISMA method, and the number of articles published in the academic field and the number of smart city pilot projects were found to be increasing in the last few years. This growth pattern also was in line with the evolution of new technologies combined with the development and exploration of traditional industries. Moreover, the importance of algorithms, especially artificial intelligence algorithms, in this process can be reflected by combining the definition and use of smart devices. Therefore, how to develop efficient and secure algorithms is another hot research area that requires more attention. The importance of smart devices and algorithms is reflected by the fact that they are relevant to each of the influencing factors. Whether it is infrastructure, or water management, or health education, they are all relevant. The acquired influences and challenges will be investigated and analyzed later to determine the case of adoption.

## 4. Research Methodology

This section discusses the research methodology used in the study. We applied the Decision-Making Trial and Evaluation Laboratory (DEMATEL) method to analyze and to identify the interrelationships between the challenges that are hindering the adoption of AI and the IoT in sustainable smart city development. DEMATEL was first employed in 1976, and it has managed to solve many global complex problems by considering experts' attitudes, and has become a widespread technique in decision-making problems [321,322]. DEMATEL is widely used in analyzing the barriers, challenges, and enablers in multiple disciplines. Recent studies have used DEMATEL for analyzing barriers in smart sustainable buildings [323], adoption of IoT [324], online food consumption [325,326], and supply-chain risk management [322,327]. The above literature emphasized the importance of DEMATEL in analyzing relationships between the challenges in various industries [61,328–331].

Figure 6 shows the various steps of the proposed methodology. There were three stages in the methodology. In stage 1, challenges for adoption of AI and IoT adoption in smart city development were identified using an extensive literature review according to the PRISMA method; in stage 2, industry experts having knowledge about the smart cities were contacted for collecting information about the challenges; and in stage 3, the collected data were analyzed, and the inter-relationships and ranking of the challenges are presented using the DEMATEL technique. Table 3 presents the profile of industry experts contacted in this study. A total of 10 experts from industries, government organizations, and universities were consulted for the data collection process. The data collection sheet was prepared in Microsoft Excel, and printed copies of the sheet were supplied to the experts. In some cases, the data sheet was sent over email to obtain the responses. The data analysis was done on a Windows-operated computer with an Intel i5 8th-generation processor with 8 GB of RAM and a 1 TB hard disk. The experts' selection was done based on a convenience sampling method. Experts with a bachelor's degree and at least two years of experience in domain knowledge of the environment, architecture, and IT roles in government and private organizations were considered. Experts with research publications and research interests in AI, smart cities, IoT technologies, and a willingness to share information were given preference. The data was collected from July 2021–August 2021.

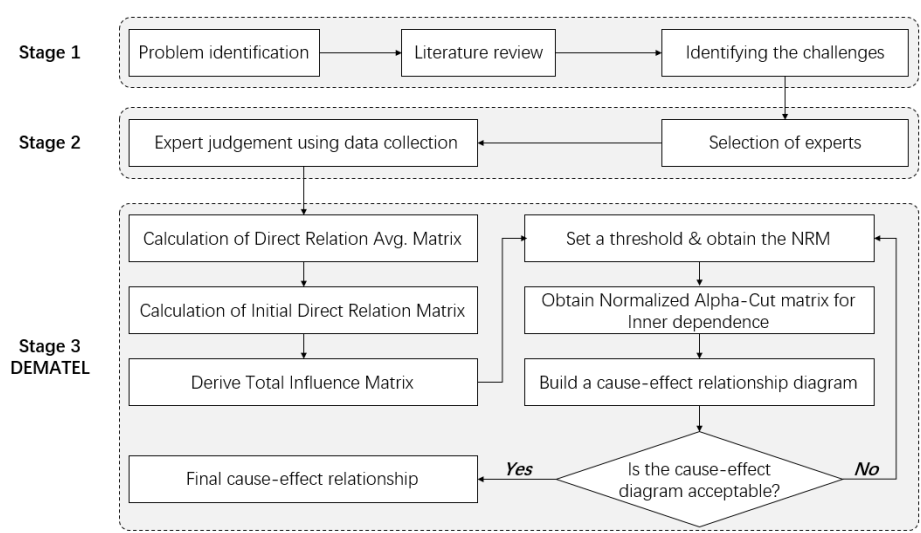

**Figure 6.** Roadmap of the research methodology.

**Table 3.** The linguistics evaluation for the assessments of the responders.

| No. | Years of Experience | Education Level | Position in Organization | Type of Industry |
|---|---|---|---|---|
| 1 | 6 | PhD in Architecture | Associate Professor | University |
| 2 | 4 | Master of Education | Senior Lecturer | University |
| 3 | 9 | Master of Science | Start-up Founder | Technical Service |
| 4 | 11 | Bachelor of Economics | Analyst Manager | Business Consultancy |
| 5 | 5 | Bachelor of Information Systems | Technical Manager | IT |
| 6 | 3 | Master of Engineering | Intern | Construction |
| 7 | 12 | Master of Engineering | Cost Supervisor | Construction |
| 8 | 7 | Bachelor of Science in Environmental Studies | Administrator | Research Institute |
| 9 | 5 | Bachelor of Art | Senior Architect | Architecture |
| 10 | 2 | Bachelor of Engineering | Planner | Government |

*Various Steps in DEMATEL Methodology (Stage 3)*

In the present study, 10 key challenges, including "Lack of infrastructure (C1)", "Insufficient funds or capital (C2)", "Cybersecurity and data risks (C3)", "Smart waste and hygiene management (C4)", "Lack of professionals (C5)", "Managing energy demands (C6)", "Managing transportation (C7)", "Environmental risks (C8)", "Managing public health and education (C9)", and "Lack of trust in AI and IoT (C10)", in the hindering of AI and IoT in the evolution of long-lasting smart cities were pointed out by reviewing substantial literature, as depicted in Table 2. A DEMATEL method (see Figure 5) was utilized to establish and analyze the causal linkages between the challenges.

We approached 10 industry experts (see Table 1 for profile details) to obtain information regarding the challenges to AI and IoT adoption for smart city development, and created a direct-relation matrix, M1, using the experts' inputs and computing it according to the formula presented in Equation (1). The causality effect of one variable (challenge) on the other variable was collected using the matrix provided in Equation (1). For our convenience, the words, "challenge" and "variable" were interchangeable in the study.

$$A_P = \begin{bmatrix} 0 & a12 & a13 & a14 & a15 & \ldots & a1n \\ a21 & 0 & a23 & a24 & a25 & \ldots & a2n \\ a31 & a32 & 0 & a34 & a35 & \ldots & a3n \\ \ldots & \ldots & \ldots & \ldots & \ldots & \ldots & \ldots \\ a(n-1)1 & a(n-2)2 & a(n-3)3 & \ldots & \ldots & 0 & a(n-1)n \\ an1 & an2 & an3 & \ldots & \ldots & an(n-1) & 0 \end{bmatrix} \quad (1)$$

where $A_P$ represents the complete matrix information collected from one expert, N represents the number of challenges, and P represents the number of experts (10) contacted for the study.

Based on the scale provided in Table 4, the experts were asked to provide their recommendations in the given format. A five-point scale running from 0–4 was used by earlier studies [316,317] for analyzing the barriers using DEMATEL. This scale ran from 0 to 4 (0—no influence, 1—very low influence, 2—low influence, 3—high influence, and 4—very high influence), as depicted in Table 4. We computed the pairwise effect of each challenge with the other using this influence scale. For example, to determine the influence between "Insufficient funds or capital (C2)" on "Lack of infrastructure (C1)" all 10 of the experts' inputs from the data collection table were extracted, and the values were 4, 1, 0, 4, 1, 4, 3, 4, 4, and 4, respectively. When computing an arithmetic mean of all experts' opinions, we get $\frac{4+1+0+4+1+4+3+4+4+4}{10} = \frac{29}{10} = 2.9$, and this value was confined to a cell (1, 2) of a direct relation matrix (DRM), which is represented in Table 5. The value 2.9 indicated a significant effect of lack of funds on lack of infrastructure. Similarly, the influence between "Managing transportation (C7)" and "Insufficient funds or capital (C2)" could be calculated as $\frac{4+4+4+4+1+3+4+4+4+4}{10} = \frac{36}{10} = 3.6$, and was confined to a cell (2, 7) of the direct relationship matrix in Table 5, which also implied a significant effect of transportation management and the availability of funds. For another computation that showed the influence between "Managing public health and education (C9)" and "Insufficient funds or capital (C2)", we get $\frac{3+0+0+3+3+3+1+0+0+0}{10} = \frac{13}{10} = 1.3$, which was specified in cell (2, 9) of DRM, and indicated a low or moderate relationship between the variables. As the same challenge could not have an effect on its own, all diagonal elements in the DRM were set to zero. Using the above procedure, all cells of DRM were calculated, and the results are shown in Table 5.

**Table 4.** The scale used for data collection for DEMATEL.

| Please Put a Suitable Value in the Matrix Based on Your Experience | Score |
|---|---|
| No Influence | 0 |
| Low Influence | 1 |
| Moderate Influence | 2 |
| High Influence | 3 |
| Very High Influence | 4 |

Step 1: Formation of a direct relationship matrix (DRM).

**Table 5.** The direct relationship matrix (M1).

| Challenges | C1 | C2 | C3 | C4 | C5 | C6 | C7 | C8 | C9 | C10 |
|---|---|---|---|---|---|---|---|---|---|---|
| C1 | 0 | 2.6 | 2.3 | 2.6 | 2.5 | 2.7 | 2.6 | 2.8 | 2.7 | 2.8 |
| C2 | 2.9 | 0 | 2 | 2.1 | 3.1 | 3 | 2.7 | 3.3 | 2.3 | 2.2 |
| C3 | 1.6 | 2.3 | 0 | 2.6 | 2.5 | 2.4 | 1.9 | 1.9 | 2.2 | 2.1 |
| C4 | 2.1 | 2.7 | 1.6 | 0 | 2.7 | 1.9 | 2.1 | 3 | 2.9 | 2 |
| C5 | 2.6 | 1.9 | 2 | 2.8 | 0 | 3 | 2.9 | 2.6 | 2.8 | 2.2 |
| C6 | 2.3 | 2.5 | 1.4 | 2.2 | 2.6 | 0 | 2.7 | 2.5 | 2.3 | 2 |
| C7 | 2.1 | 3.6 | 2.1 | 2.7 | 2 | 2.9 | 0 | 2.3 | 2.5 | 1.9 |
| C8 | 2.1 | 2.5 | 2.3 | 1.9 | 2.5 | 2.7 | 2.8 | 0 | 2.5 | 1.8 |
| C9 | 2.5 | 1.3 | 1.9 | 2.6 | 2.8 | 2.5 | 2.5 | 2.3 | 0 | 2.8 |
| C10 | 1.8 | 1.9 | 3.2 | 2.8 | 2.2 | 2.2 | 2.8 | 3.4 | 3 | 0 |

Step 2: Formation of a normalized direct relation matrix (NRM).

To prepare the data for the DEMATEL analysis, an NRM was prepared. The normalized direct-relation matrix "M2" (see Table 6) was constructed using the data based on the original direct-relation matrix. Each cell was divided by the sum of all cell values in that corresponding row. For example, for the DRM in row 1, the summation of all values (2.6, 2.3, 2.6, 2.5, 2.7, 2.6, 2.8, 2.7, and 2.8) equaled 23.6. Each cell in row 1 of Table 5 was divided by this corresponding row total of 23.6. Then, the normalized values were obtained. Similarly, for all rows of the DRM, the normalization is done to obtain the NRM, which is shown in Table 6.

**Table 6.** The normalized direct-relation matrix (M2).

| Challenges | C1 | C2 | C3 | C4 | C5 | C6 | C7 | C8 | C9 | C10 |
|---|---|---|---|---|---|---|---|---|---|---|
| C1 | 0 | 0.1079 | 0.0954 | 0.1079 | 0.1037 | 0.112 | 0.1079 | 0.1162 | 0.112 | 0.1162 |
| C2 | 0.1203 | 0 | 0.083 | 0.0871 | 0.1286 | 0.1245 | 0.112 | 0.1369 | 0.0954 | 0.0913 |
| C3 | 0.0664 | 0.0954 | 0 | 0.1079 | 0.1037 | 0.0996 | 0.0788 | 0.0788 | 0.0913 | 0.0871 |
| C4 | 0.0871 | 0.112 | 0.0664 | 0 | 0.112 | 0.0788 | 0.0871 | 0.1245 | 0.1203 | 0.083 |
| C5 | 0.1079 | 0.0788 | 0.083 | 0.1162 | 0 | 0.1245 | 0.1203 | 0.1079 | 0.1162 | 0.0913 |
| C6 | 0.0954 | 0.1037 | 0.0581 | 0.0913 | 0.1079 | 0 | 0.112 | 0.1037 | 0.0954 | 0.083 |
| C7 | 0.0871 | 0.1494 | 0.0871 | 0.112 | 0.083 | 0.1203 | 0 | 0.0954 | 0.1037 | 0.0788 |
| C8 | 0.0871 | 0.1037 | 0.0954 | 0.0788 | 0.1037 | 0.112 | 0.1162 | 0 | 0.1037 | 0.0747 |
| C9 | 0.1037 | 0.0539 | 0.0788 | 0.1079 | 0.1162 | 0.1037 | 0.1037 | 0.0954 | 0 | 0.1162 |
| C10 | 0.0747 | 0.0788 | 0.1328 | 0.1162 | 0.0913 | 0.0913 | 0.1162 | 0.1411 | 0.1245 | 0 |

Step 3: Calculation of the total relation matrix (TRM).

Further, by taking into account the normalized direct-relation matrix (M2), a total-relation matrix (M3) was calculated by using the following equation:

$$T = X (I - X)^{-1} \qquad (2)$$

where *T* represents the notation for TRM, X denotes NRM, and I denotes the identity matrix. Then using the procedure mentioned in [222], threshold values ($\alpha$), D, and R were calculated, where D represents the sum of the values in a row, and R represents the sum of values in a column. Equations (3)–(5) were used to calculate the values D, R, $\alpha$ values:

$$D = [d_{i_i}]_{n \times 1} = [\sum_{j=1}^{n} d_{ij}]_{n*1} \qquad (3)$$

$$R = [r]_{1 \times n} = [\sum_{i=1}^{n} r_{ij}]_{1*n} \qquad (4)$$

$$\alpha = \frac{\sum_{j=1}^{n} \cdot \sum_{i=1}^{n} r_{i_j}}{n^2} \tag{5}$$

where i and j are the respective values for the rows and columns in the TRM, and n is the number of challenges in the study. The threshold values were used to determine the important and inconsequential challenges based on the threshold value. The $\alpha$ score was obtained as 0.9752, and the cell values in the TRM matrix that were smaller than the threshold $\alpha$ value (0.9752) were replaced with the value "zero (0)" in the $\alpha$-cut TRM (see Table 7), and were ignored for further DEMATEL processing. The $\alpha$ value of challenges, which was equal to or greater than the threshold value, was updated in a new matrix called $\alpha$- cut TRM (Table 8).

**Table 7.** The total direct-relation matrix (M3).

| Challenges | C1 | C2 | C3 | C4 | C5 | C6 | C7 | C8 | C9 | C10 |
|---|---|---|---|---|---|---|---|---|---|---|
| C1 | 0.8878 | 1.0343 | 0.916 | 1.0695 | 1.0936 | 1.1181 | 1.1042 | 1.1496 | 1.1117 | 0.977 |
| C2 | 0.9975 | 0.939 | 0.9062 | 1.0534 | 1.1153 | 1.131 | 1.1099 | 1.1675 | 1.0992 | 0.9574 |
| C3 | 0.8074 | 0.872 | 0.6939 | 0.9126 | 0.9325 | 0.9431 | 0.9169 | 0.9496 | 1.0925 | 0.9374 |
| C4 | 0.8815 | 0.9441 | 0.8093 | 0.8755 | 1.0023 | 0.9909 | 0.988 | 1.0532 | 1.0186 | 0.8634 |
| C5 | 0.956 | 0.9802 | 0.9917 | 1.0444 | 0.9663 | 1.0946 | 1.081 | 1.1074 | 1.0818 | 0.9276 |
| C6 | 0.8721 | 0.9221 | 0.7871 | 0.9419 | 0.98 | 0.8992 | 0.9905 | 1.0172 | 0.9797 | 0.8472 |
| C7 | 0.9182 | 1.0153 | 0.8595 | 1.0159 | 1.0201 | 1.0668 | 0.9485 | 1.0727 | 1.0457 | 0.8954 |
| C8 | 0.8814 | 0.9396 | 0.8334 | 0.9499 | 0.9954 | 1.0192 | 1.0121 | 0.9415 | 1.0047 | 0.8566 |
| C9 | 0.8984 | 0.9023 | 0.8251 | 0.98 | 1.0097 | 1.0159 | 1.007 | 1.0348 | 0.9176 | 0.8959 |
| C10 | 0.939 | 0.9921 | 0.9308 | 1.0574 | 1.0635 | 1.0805 | 1.0901 | 1.1476 | 1.1018 | 0.8549 |

**Table 8.** The $\alpha$-cut total direct-relation matrix (M4).

| Challenges | C1 | C2 | C3 | C4 | C5 | C6 | C7 | C8 | C9 | C10 |
|---|---|---|---|---|---|---|---|---|---|---|
| C1 | 0 | 1.0343 | 0 | 1.0695 | 1.0936 | 1.1181 | 1.1042 | 1.1496 | 1.1117 | 0.977 |
| C2 | 0.9975 | 0 | 0 | 1.0534 | 1.1153 | 1.131 | 1.1099 | 1.1675 | 1.0992 | 0 |
| C3 | 0 | 0 | 0 | 0 | 0 | 0 | 0 | 0.1614 | 0.1259 | 0 |
| C4 | 0 | 0 | 0 | 0 | 1.0023 | 0.9909 | 0.988 | 1.0532 | 1.0186 | 0 |
| C5 | 0 | 0.9802 | 0.1147 | 1.0444 | 0 | 1.0946 | 1.081 | 1.1074 | 1.0818 | 0 |
| C6 | 0 | 0 | 0 | 0 | 0.98 | 0 | 0.9905 | 1.0172 | 0.9797 | 0 |
| C7 | 0 | 1.0153 | 0 | 1.0159 | 1.0201 | 1.0668 | 0 | 1.0727 | 1.0457 | 0 |
| C8 | 0 | 0 | 0 | 0 | 0.9954 | 1.0192 | 1.0121 | 0 | 1.0047 | 0 |
| C9 | 0 | 0 | 0 | 0.98 | 1.0097 | 1.0159 | 1.007 | 1.0348 | 0 | 0 |
| C10 | 0 | 0.9921 | 0 | 1.0574 | 1.0635 | 1.0805 | 1.0901 | 1.1476 | 1.1018 | 0 |

The $\alpha$-cut total DRM was prepared by removing the matrix values less than the threshold value (0.9752), all the values that were less than $\alpha$ were assigned a 0 value, and all the values other than zero shown in Table 8 were considered to build the directed graph (digraph). The arrows in the digraph (see Figure 7) indicate the relationships between the challenges.

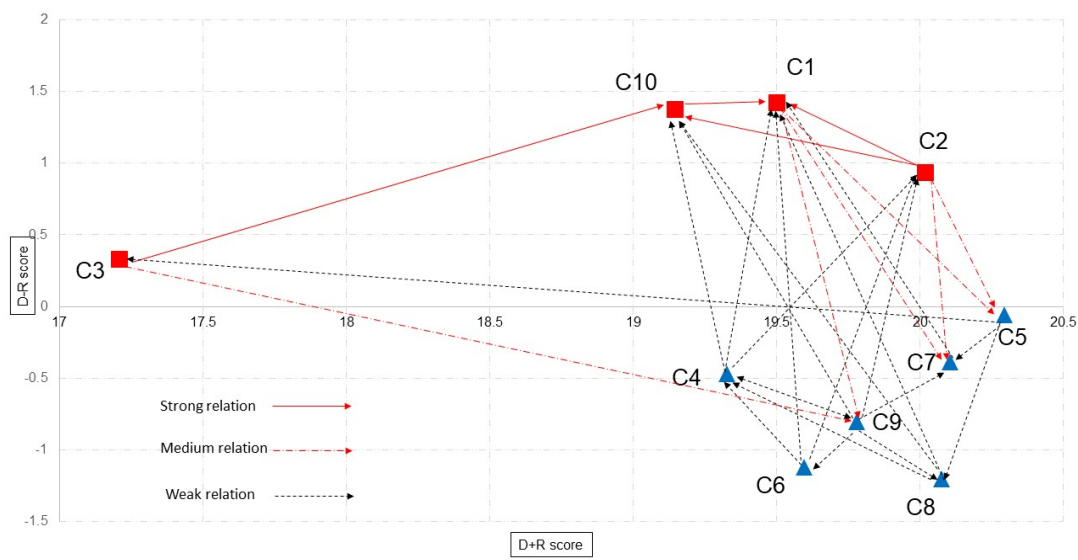

**Figure 7.** Directed graph showing the relationships between the challenges.

. For example, the cybersecurity risks category is in the rightmost corner of the map, and it has only one arrow pointing towards it, which indicates that the challenge "lack of professionals (C5)" influenced "cybersecurity and data risks (C3)". In addition, "cybersecurity and data risk (C3)" has two outgoing arrows towards "managing public health and education (C9)" and "lack of trust in AI and IoT (C10)". The overall network graph analysis concluded that the other challenges had much less influence on the cybersecurity risks. The bidirectional arrows represent the dual relationship between the given set of challenges. The findings of R and D corroborated the degree of relationship effect among each critical challenge. While D + R represents the significance of a particular challenge, D-R shows the net influence of the provided challenge. The D + R score is plotted on the *X*-axis, and the D-R score is plotted on the *Y*-axis to plot the cause–effect challenges on the scatter plot. For instance, for computations of D + R and D-R for challenge C1; the score of D was 10.462 and the score of R was 9.0393, so adding them together (D + R) equaled 19.50125, whereas subtracting them (D-R) equaled 1.4226752. The challenges with high D-R values had a high importance, and were named as the "cause" group or causal challenges, and the challenges with low D-R scores were less significant and influenced by the cause group challenges, and were named as the "effect" group challenges. All results are presented in Table 9.

Table 9. Prominence and relation results obtained by using the DEMATEL method.

| Criteria Name | Code | D Score | R Score | D + R Score | D-R Score | Type |
|---|---|---|---|---|---|---|
| Lack of infrastructure | C1 | 10.462 | 9.0393 | 19.50125 | 1.4226752 | Cause |
| Insufficient funds or capital | C2 | 10.477 | 9.5411 | 20.01761 | 0.9353911 | Cause |
| Cybersecurity risks | C3 | 8.7706 | 8.4383 | 17.20887 | 0.3322851 | Cause |
| Smart Waste and hygiene management | C4 | 9.4267 | 9.9004 | 19.32717 | −0.473676 | Effect |
| Lack of professionals | C5 | 10.116 | 10.179 | 20.29484 | −0.062452 | Effect |
| Managing energy demands | C6 | 9.2369 | 10.359 | 19.59631 | −1.122412 | Effect |
| Managing transportation | C7 | 9.8581 | 10.248 | 20.1064 | −0.390191 | Effect |
| Environmental risks | C8 | 9.434 | 10.641 | 20.07508 | −1.207166 | Effect |
| Managing public health and education | C9 | 9.4868 | 10.292 | 19.77877 | −0.805249 | Effect |
| Lack of trust in AI, IoT | C10 | 10.258 | 8.887 | 19.14481 | 1.3707933 | Cause |

As shown in Table 9, the barrier with a D–R value less than zero was identified as an effective group, while a barrier with more than the D–R value fell under the cause group. Based on the DEMATEL results as shown in Table 9 and Figure 8, the causal interactions and the degrees of influence among the AI and IoT adoption challenges in the smart city development are explained as follows. The challenges "C1, C2, C3, C10" were the main challenges affecting the implementation of various challenges for adoption of AI and the IoT in sustainable smart city development. Figure 8 displays the differentiation between the cause and effect groups. The challenges above the horizontal axis fell under the "cause" group, while the challenges below the horizontal line fell under the "effect" group.

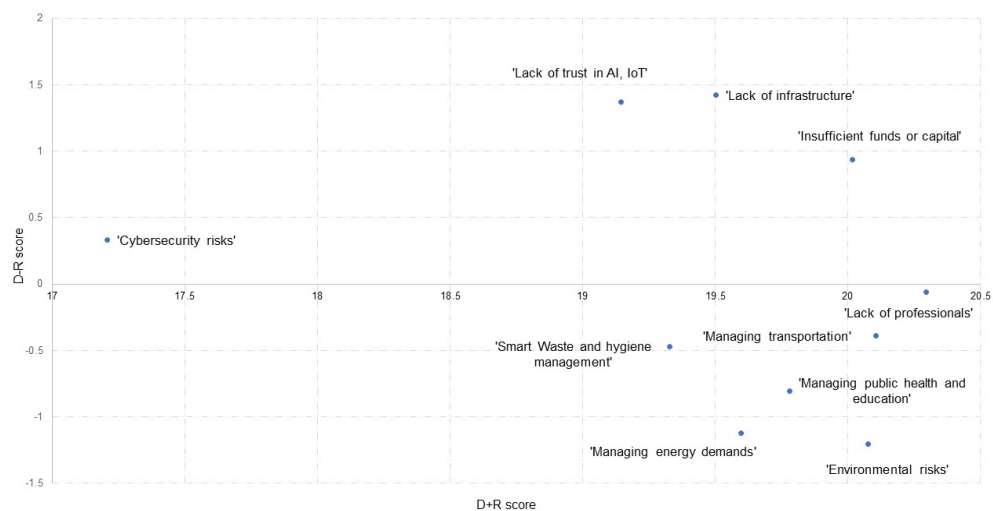

**Figure 8.** The cause–effect diagram.

Figure 7 displays the relationship between the cause–effect challenges; whereas the red colored solid lines show the strong relationship between the challenges, the red colored dashed lines indicate a moderate relationship, and the black colored dotted lines show the weak relationships. Researchers and organizations should assign prime importance to strong relationships, as understanding the connection between the challenges will help to overcome issues associated with the adoption of AI and the IoT in smart city development. To understand the relationships between the various challenges, we used NRM (see Figure 7), which displayed the weak, moderate, and strong causal linkages between all 10 challenges. The cause–effect relationships and their importance is explained in Section 5.

## 5. Results and Discussion

The following section summarizes the findings based on the causal diagram (see Figures 6 and 7). As shown in Table 9, the challenges with a D–R value less than zero were identified as the "effect" group, while challenges with more than the D–R value fell under the "cause" group. The challenges, including Lack of infrastructure (C1), Insufficient funds or capital (C2), Cybersecurity risks (C3), and Lack of trust in AI, IoT (C10) were classified into the cause criteria group, while affecting the remaining challenges, which included Smart Waste and hygiene management (C4), Lack of professionals (C5), Managing energy demands (C6), Managing transportation (C7), Environmental risks (C8), and Managing public health and education (C9). Because cause variables have an impact on the effect group criterion, they should be given more attention while planning AI and IoT technologies in smart city development. The effect group challenges, although they had less dominance over the other challenges, were highly influenced by the cause group challenges. As mentioned in [328,329], cause group factors will have a strong relationship and high prominence. The factors that fall under the cause group are driving factors. The factors directly linked with cause group factors will have a high influence on these variables, and they have a strong relationship and low prominence. The factors (barriers) that are not connected to other factors will have no influence on or prominence in other factors.

The most important challenge in adopting AI and the IoT to achieve a large competitive factor in a sustainable smart city was "Insufficient funds or capital (C2)", with the greatest D-R score of 1.422, implying that (C2) should be given greater emphasis in the entire system of IoT and AI implementation when designing an advanced city. This finding held true because recent studies have identified the unavailability of funds or financial resources as a major constraint for smart city transformation projects [69,332]. Furthermore, Table 9 reveals that the important effect degree of (C1) was 10.462, which ranked second-highest among all causative factors. In general, (C1) was a major element that requires further consideration during the implementation phase, and this finding also was supported by the literature [333]. With the second-greatest D-R value, "Lack of trust in IoT and AI (C10)" had a substantial influence on the other challenges. Lack of trust creates a barrier for building trust and promoting the use of AI and the IoT in smart city projects. One more important challenge identified from the analysis was "Cybersecurity risks (C3)", with a D-R value of 0.3352, which also held in the context of smart city development. The interconnection and speed of the Internet and availability of vast public data makes smart cities vulnerable to cyber security attacks. An example of such a cyber security attack in the USA [334] was discussed in the introduction section of this study.

If the value of D-R was negative, the perspective belonged to the impact group (effects), and they were heavily influenced by cause group challenges. Furthermore, in terms of notable effect degree, (C6) and (C8) were highly influenced by the cause group challenges. Similarly, the remaining challenges C4, C7, and C9 had low D-R values, which implied a significant low importance of these challenges in the smart city development in the context of AI and IoT adoption. In this study, "Environmental risks (C8)" had an R value of 10.641, the highest out of all challenges. Furthermore, when contrasted with other variables in the effect group, its D-R value was quite high (−1.2071). To understanding this trade-off clearly, a cause–effect diagram was used. In Figure 8, it can be observed that the challenge regarding environmental risks fell on the right side of the graph, which implied that even though the C8 challenge fell under the effect group, it had a substantial influence on the other elements. The D + R for "Managing transportation (C7)" was the second-largest in the entire procedure. Its D-R value was quite low, but "Managing energy demand (C6)" and "Managing public health and education (C9)" had the second- and third-highest R values of 10.359 and 10.292, respectively, as well as the second- and third-highest D-R values of −1.122412 and −0.8052, respectively. Both also can be considered on a priority basis. However, their D+R scores were lower than the other effect group criteria. The other components had mediated R values, indicating a high degree of effect. "Lack of professionals (C5)" had a negative influence on the system, since its D-R value was −0.0624.

Other challenges had a major influence on it. The digraph of net cause and effect is shown in Figure 7, and it was used to demonstrate the relationship.

In addition, based on the M4 matrix results displayed in Table 6, we created a causal inter-relationship graph of the implementation of AI and IoT challenges in the development of the smart city and the interaction among these; key challenges are shown in Figure 7. The arrow-headed lines indicate the causal interactions among each pair of challenges. The red-colored labels are causal challenges, and the blue-colored labels represent the effect group challenges. It was found that "Lack of professional (C5)" had strong interactions with another challenge, "Managing energy demands (C6)", while "Managing transportation (C7)", "Environmental risks (C8)", and "Managing public health and education (C9)" were further related by highly influencing, or having more interactions with, other challenges to the implementation of the IoT and AI in the development of a smart city (see Table 9).

According to this study, "Lack of infrastructure (C1)", "Insufficient funds (C2)", "Cyber security risks (C3)", "Lack of trust on AI, IoT models (C10)" were highly dominant challenges that affected the other challenges. The organizations that are planning to implement AI and the IoT in smart city development should focus on these challenges first, as overcoming these challenges will facilitate the adoption process. The findings of the study will help to identify important challenges, and the relationship and influence of challenges also can be extracted. As a result, researchers and policymakers researching sustainable smart city development must address the major challenges to implementing IoT and AI.

## 6. Conclusion and Future Discussion

The concept of smart cities is becoming important to providing sustainable living conditions to people. Achieving sustainability is difficult due to the involvement of multiple factors. AI and IoT technologies can transform challenges and offer potential solutions to the issues perceived by society and regulatory bodies. This study aimed to identify and analyze the effect of important challenges that are hindering the adoption of AI and the IoT in sustainable smart city development. Initially, the challenges were extracted from an extensive literature review, then the DEMATEL method was used to analyze the relationship of causes–effects between the identified challenges. Data were collected from experts in diverse industries familiar with smart cities. Due to the rapid urbanization worldwide, cities are becoming overcrowded and are facing issues such as water scarcity, pollution, soil contamination, etc. On the other hand, unavailability of funds; global warming; and providing food security, basic health, and education are the key issues facing government bodies. Previous studies elaborated that AI and IoT technologies were useful in analyzing the optimization of resources using accurate data collection processes. Developing plans for sustainable smart cities is a major challenge for emerging economies. Some countries (e.g., Singapore, Denmark, etc.) have leveraged the benefits of the AI and IoT techniques to measure and track various activities such as pollution levels, weather monitoring, energy distribution and management, traffic and transport management, water distribution and sewerage monitoring, etc. All the data collected by various sensors, cameras, videos can be sent to a central platform to perform an analysis to identify the issues and possible solutions using analytics. For example, the benefits of AI in traffic management to improve pedestrian safety is followed in Singapore [335], and Copenhagen uses an AI-enabled solutions lab to optimize energy usage and weather monitoring [335]. However, the majority of the smart city development is occurring in developed countries, whereas emerging economies such as China, India, and Brazil are still struggling to overcome the challenges in developing smart cities. The huge population and rapid urbanization rates are causing these emerging economies to design and develop sustainable smart cities at the present time.

Using the data from experts' opinions, the identified challenges were analyzed using the DEMATEL method. The challenges were grouped into cause types and effect types based on their importance as determined by the experts. Using the findings, we aimed to explore the relationships between the challenges and how they influenced the adoption of AI and the IoT in sustainable smart city development. Among all challenges, Lack of infrastructure (C1), Insufficient funds or capital (C2), Cybersecurity risks (C3), and Lack of trust in AI, IoT (C10) were the most prominent factors influencing the adoption of AI and the IoT in smart city development. Therefore, researchers and regulatory bodies should give importance to these challenges while planning a smart city project. In the causal loop diagram, it was observed that Smart waste and hygiene management (C4), Lack of professionals (C5), Managing energy demand (C6), Managing transportation (C7), Environmental risks (C8), and Managing public health and education (C9) were highly affected by the other factors. The relationships influencing these challenges can help us to understand the limitations and uncertainties associated with the other challenges. The results of this study can help academicians, industry practitioners, and policymakers reach an overall understanding of the inter-relationships between the challenges that are hindering the adoption of AI and the IoT in sustainable smart city development. The data were collected from experts from diverse industries and organizations, which made this study unique for emerging economies. A summary of the highlights of this study follows.

- We identified challenges for adopting AI and the IoT for sustainable smart city development using a systematic literature review method (PRISMA);
- Ten key challenges were identified using the PRISMA method, and a total of 171 articles were studied;
- Ten experts having experience in smart city development were contacted, and their opinions were evaluated using DEMATEL;
- The study findings revealed that lack of infrastructure, lack of trust in AI and the IoT, cyber security and data risks, and insufficient funds were the major barriers that are affecting the adoption of AI and the IoT for sustainable smart city development;
- The results of this study can be used as a basis for understanding various inter-relationships between the barriers, as the study was based on data from China and developing countries such as India, Brazil, and Malaysia can observe the findings as useful while adopting AI and the IoT for smart city projects.

Despite many efforts taken to carefully design the study, there were some shortcomings. We did not use a Grey scale to measure the accuracy of opinions of experts about the challenges, so in future research, Grey-DEMATEL with a greater number of challenges can be explored to achieve good results in this domain. ANP, AHP, or ISM can be used to present the hierarchical interrelationships more robustly. In addition, the consultation with experts with great experience in this domain can help reach possible solutions to overcome the challenges hindering sustainable smart city development.

**Author Contributions:** Conceptualization, K.W.; methodology, R.K.G.; software, R.K.G.; validation, K.W., Y.Z., and R.K.G.; formal analysis, K.W.; investigation, K.W. and Z.L.; resources, K.W.; data curation, K.W. and Z.L.; writing—original draft preparation, R.K.G. and Z.L.; writing—review and editing, K.W. and R.K.G.; visualization and formatting, Z.L.; supervision, K.W.; project administration, Y.Z.; funding acquisition, Y.Z. K.W., Y.Z., and Z.L. contributed equally to this work. All authors have read and agreed to the published version of the manuscript.

**Funding:** This research and the APC were funded by the University-Enterprise-Partnership Program of Solearth Architecture (grant number DOU- 315324C-KGT).

**Institutional Review Board Statement:** Not applicable.

**Informed Consent Statement:** Not applicable.

**Data Availability Statement:**  Not applicable.

**Conflicts of Interest:** The authors declare no conflict of interest.

**References**

1. Wu, X.; Yang, Z. Smart City Concept and Future City Development. *Urban Dev. Res.* **2010**, *11*, 60.
2. Palmisano, S.J. A Smarter Planet: The next Leadership Agenda. *IBM* **2008**, *6*, 1–8.
3. Fernandez-Anez, V. Stakeholders Approach to Smart Cities: A Survey on Smart City Definitions. In *Proceedings of the International Conference on Smart Cities*; Springer: Berlin/Heidelberg, Germany, 2016; pp. 157–167.
4. Gracia, T.J.H.; García, A.C. Sustainable Smart Cities. Creating Spaces for Technological, Social and Business Development. *Boletín Científico Cienc. Económico Adm. ICEA* **2018**, *6*, 3074, doi.org/10.29057/icea.v6i12.3074.
5. Lee, S.K.; Kwon, H.R.; Cho, H. *International Case Studies of Smart Cities: Singapore*; Inter-American Development Bank: New York, NY, USA, 2016;.
6. Khan, H.H.; Malik, M.N.; Zafar, R.; Goni, F.A.; Chofreh, A.G.; Klemeš, J.J.; Alotaibi, Y. Challenges for sustainable smart city development: A conceptual framework. *Sustain. Dev.* **2020**, *28*, 1507–1518.
7. Treude, Mona. Sustainable Smart City—Opening a Black Box. *Sustainability* **2021**, *13*, 769.
8. Kim, T.; Ramos, C.; Mohammed, S. Smart City and IoT. *Future Gener. Comput. Syst.* **2017**, *76*, 159–162, doi:10.1016/j.future.2017.03.034.
9. UN DESA. *World's Population Increasingly Urban with More than Half Living in Urban Areas*; United Nations Department of Economic and Social Affairs: New York, NY, USA, 2014.
10. Yang, D.; Li, P. The Economic Effects of Population Agglomeration: An Empirical Study Based on Instrumental Variables. *J. Demogr.* **2019**, *3*, 28–37.
11. Mohurd.gov.cn. Ministry of Housing and Urban-Rural Development of the People's Republic of China-Notice of the General Office of the Ministry of Housing and Urban-Rural Development on Announcement of the List of National Smart City Pilots in 2013. 2021. Available online: http://www.mohurd.gov.cn/wjfb/201308/t20130805_214634.html (accessed on 20 September 2021).
12. NDRC. China's New Smart City Development Status, Situation and Policy Recommendations. 2021. Available online: https://www.ndrc.gov.cn/xxgk/jd/wsdwhfz/202005/t20200515_1228150.html?code=&state=123 (accessed on 20 September 2021).
13. Mohurd.gov.cn. Ministry of Housing and Urban-Rural Development of the People's Republic of China-97 Cities were Selected into the National Smart City Pilot List and 41 Projects Were Identified as Special Pilots. 2021. Available online: http://www.mohurd.gov.cn/zxydt/201504/t20150414_220664.html (accessed on 20 September 2021).
14. Mohurd.gov.cn. Ministry of Housing and Urban-Rural Development of the People's Republic of China-Notice of the General Office of the Ministry of Housing and Urban-Rural Development and the General Office of the Ministry of Science and Technology on Announcement of the National Smart City Pilot List in 2014. 2021. Available online: http://www.mohurd.gov.cn/wjfb/201504/t20150410_220653.html (accessed on 20 September 2021).
15. Lu, Z. Foresight 2021: "Panorama of China's Smart City Construction Industry in 2021" (with Market Status, Development Trend, etc.). Qianzhan. 2021. Available online: https://www.qianzhan.com/analyst/detail/220/210208-0761c450.html (accessed on 20 September 2021]=).
16. Qianzhan. Analysis of the overall situation of China's pilot smart cities in many places to speed up construction planning. 2021. Available online: https://f.qianzhan.com/chanyeguihua/detail/200618-2af45f4a.html (accessed on 20 September 2021).
17. Qianzhan. 2019 China Smart City Industry Market Status and Development Prospects Analysis Forecast to Reach 25 Trillion by 2022. 2019. Available online: https://bg.qianzhan.com/trends/detail/506/191206-35e306f4.html (accessed on 20 September 2021).
18. Qianzhan. China's Smart City Industry Market Status and Development Prospects in 2020 the Market Size Will Reach about 25 Trillion Yuan in 2022. 2020. Available online: https://bg.qianzhan.com/report/detail/300/200722-86ed4f0d.html (accessed on 20 September 2021.
19. Appleton, J. How Smart Cities Are Boosting Citizen Engagement. Hub.beesmart.city. 2021. Available online: https://hub.beesmart.city/en/strategy/how-smart-cities-boost-citizen-engagement (accessed on4 September 2021).
20. Haas, K. How To Build A Smart City: Decision Making and Modeling. Skyfii|Omnidata Intelligence Solutions for Physical Venues. 2021. Available online: https://skyfii.io/blog/how-to-build-a-smart-city-decision-making-and-modeling/ (accessed on 4 September 2021).
21. ANBOUND Research Center. How to Design and Operate Smart Cities?. Available online: https://www.anbound.my/Section/ArticalView.php?Rnumber=18377&SectionID=1 (accessed on 1 Oct 2021).
22. Cohen, B. The 3 Generations of Smart Cities. Available online: https://www.fastcompany.com/3047795/the-3-generations-of-smart-cities (accessed on 17 July 2021).
23. Gao, Bai, and Yi Ru. Industrial policy and competitive advantage: A comparative study of the cloud computing industry in Hangzhou and Shenzhen. In *Innovation and China's Global Emergence*; NUS Press: Singapore, 2021; pp. 232–262.
24. Shah, J.; Kothari, J.; Doshi, N. A Survey of Smart City Infrastructure via Case Study on New York. *Procedia Comput. Sci.* **2019**, *160*, 702–705, doi:10.1016/j.procs.2019.11.024.
25. Tobias, M. How New York Is Becoming a Smart City. Available online: https://www.ny-engineers.com/blog/how-new-york-is-becoming-a-smart-city (accessed on 17 July 2021).
26. Tomás, J.P. Case Study: Smart City Technology in Singapore. Available online: https://enterpriseiotinsights.com/20170112/smart-cities/case-study-singapore-smart-city-tag23-tag99 (accessed on 17 July 2021).
27. ASKCI China's Smart City Industry Chain Of Communication Network Layer Market Development Status Analysis In 2020. Available online: https://m.askci.com/news/chanye/20200603/0922391161320.shtml (accessed on 17 July 2021).

28. Qianzhan Industry Institute Analysis of the Current Situation And Development Trend of the Industry Chain Of China's Smart City Industry In 2020. Available online: https://bg.qianzhan.com/report/detail/300/200727-97baa9b8.html (accessed on 17 July 2021).
29. Wu, X. Panoramic Combing Of Smart City Industry Chain and Regional Heat Map. Available online: https://finance.stockstar.com/IG2021050600000756.shtml (accessed on 17 July 2021).
30. Zhidongxi 2020 Smart City Industry Chain Map: Seven Links And Five Subsectors Explained In A Comprehensive Manner. Available online: https://www.163.com/dy/article/FV10UOQL0511QE18.html (accessed on 17 July 2021).
31. Zhuang, Z.; Fa, X. City Brain: Building the Golden Key to Digital Urban Governance. *Hangzhou Daily* 20 Aug **2019**; Page A11.
32. Shenzhen Special Zone Daily 2020 Is About To Close, Shenzhen Smart City Construction To Deliver A Bright Answer Sheet. Available online: https://www.sohu.com/a/www.sohu.com/a/440978762_321029 (accessed on 18 July 2021).
33. Byun, J.; Kim, S.; Sa, J.; Kim, S.; Shin, Y.-T.; Kim, J.-B. Smart City Implementation Models Based on IoT Technology. *Adv. Sci. Technol. Lett.* **2016**, *129*, 209–212.
34. Theodoridis, E.; Mylonas, G.; Chatzigiannakis, I. Developing an Iot Smart City Framework. In Proceedings of the IISA 2013, Piraeus, Greece, 10–12 July 2013; pp. 1–6.
35. Latre, S.; Leroux, P.; Coenen, T.; Braem, B.; Ballon, P.; Demeester, P. City of Things: An Integrated and Multi-Technology Testbed for IoT Smart City Experiments. In Proceedings of the 2016 IEEE International Smart Cities Conference (ISC2), Trento, Italy, 12–15 September 2016; pp. 1–8.
36. Sanchez, L.; Muñoz, L.; Galache, J.A.; Sotres, P.; Santana, J.R.; Gutierrez, V.; Ramdhany, R.; Gluhak, A.; Krco, S.; Theodoridis, E.; et al. SmartSantander: IoT Experimentation over a Smart City Testbed. *Comput. Netw.* **2014**, *61*, 217–238, doi:10.1016/j.bjp.2013.12.020.
37. Li, Q. Urban Information Visualization Design Study. Ph.D. Thesis, Shanghai University, Shanghai, China, 2017.
38. Chen, Q.; Qin, J. Architecture Design Of A Cloud Platform For Real-time Management Of Building Energy Consumption And Environment For Big Data. *Green Building*; 2019, Vol.1, pp. 77–80.
39. Zhang, Y.; Han, H.; Yang, H.; Yang, C.; Wang, Z. Data Mining Technology and its Application in Building Energy Efficiency. *Comput. Syst. Appl.* **2017**, *26*, 151–157.
40. Shi, L.; Tang, J.; Wang, X. *A Data-Based Technique for Analyzing and Predicting Building Energy Consumption (Patent No. CN107480371A)*; Shanghai, China. 2017.
41. Yang, S.; Luo, S.; Du, M.; Gu, Z.; Wu, L.; Zhong, Y.; Li, X.; Li, D.; Zhang, H. Data Processing Method Of Public Building Energy Consumption Supervision Platform Based On Data Mining. *Deying* **2015**, Vol.2, pp. 82–86.
42. Zhao, Y.; Genovese, P.V.; Li, Z. Intelligent Thermal Comfort Controlling System for Buildings Based on IoT and AI. *Future Internet* **2020**, *12*, 30.
43. Li, Z.; Genovese, P.V.; Zhao, Y. Study on Multi-Objective Optimization-Based Climate Responsive Design of Residential Building. *Algorithms* **2020**, *13*, 238.
44. Tian, Z.; Chen, W.; Shi, X.; Si, B. Integrating EnergyPlus and Dakota to Optimize Building Energy Consumption and Case Studies. *Build. Technol. Dev.* **2016**, Vol.6, pp. 73–76.
45. Amasyali, K.; El-Gohary, N.M. A Review of Data-Driven Building Energy Consumption Prediction Studies. *Renew. Sustain. Energy Rev.* **2018**, *81*, 1192–1205.
46. Sadeghi, A.; Younes Sinaki, R.; Young, W.A.; Weckman, G.R. An Intelligent Model to Predict Energy Performances of Residential Buildings Based on Deep Neural Networks. *Energies* **2020**, *13*, 571.
47. Dong, B.; Li, Z.; Rahman, S.M.; Vega, R. A Hybrid Model Approach for Forecasting Future Residential Electricity Consumption. *Energy Build.* **2016**, *117*, 341–351.
48. Li, K.; Hu, C.; Liu, G.; Xue, W. Building's Electricity Consumption Prediction Using Optimized Artificial Neural Networks and Principal Component Analysis. *Energy Build.* **2015**, *108*, 106–113.
49. Ahmad, A.S.; Hassan, M.Y.; Abdullah, M.P.; Rahman, H.A.; Hussin, F.; Abdullah, H.; Saidur, R. A Review on Applications of ANN and SVM for Building Electrical Energy Consumption Forecasting. *Renew. Sustain. Energy Rev.* **2014**, *33*, 102–109.
50. Daut, M.A.M.; Hassan, M.Y.; Abdullah, H.; Rahman, H.A.; Abdullah, M.P.; Hussin, F. Building Electrical Energy Consumption Forecasting Analysis Using Conventional and Artificial Intelligence Methods: A Review. *Renew. Sustain. Energy Rev.* **2017**, *70*, 1108–1118.
51. Turhan, C.; Kazanasmaz, T.; Uygun, I.E.; Ekmen, K.E.; Akkurt, G.G. Comparative Study of a Building Energy Performance Software (KEP-IYTE-ESS) and ANN-Based Building Heat Load Estimation. *Energy Build.* **2014**, *85*, 115–125.
52. Qiu, C.; Yi, Y.K.; Wang, M.; Yang, H. Coupling an Artificial Neuron Network Daylighting Model and Building Energy Simulation for Vacuum Photovoltaic Glazing. *Appl. Energy* **2020**, *263*, 114624.
53. Zhong, H.; Wang, J.; Jia, H.; Mu, Y.; Lv, S. Vector Field-Based Support Vector Regression for Building Energy Consumption Prediction. *Appl. Energy* **2019**, *242*, 403–414.
54. Roldán-Blay, C.; Escrivá-Escrivá, G.; Álvarez-Bel, C.; Roldán-Porta, C.; Rodríguez-García, J. Upgrade of an Artificial Neural Network Prediction Method for Electrical Consumption Forecasting Using an Hourly Temperature Curve Model. *Energy Build.* **2013**, *60*, 38–46.
55. Goyal, S.; Barooah, P.; Middelkoop, T. Experimental Study of Occupancy-Based Control of HVAC Zones. *Appl. Energy* **2015**, *140*, 75–84.

56. Williams, K.T.; Gomez, J.D. Predicting Future Monthly Residential Energy Consumption Using Building Characteristics and Climate Data: A Statistical Learning Approach. *Energy Build.* **2016**, *128*, 1–11.
57. Massana, J.; Pous, C.; Burgas, L.; Melendez, J.; Colomer, J. Short-Term Load Forecasting in a Non-Residential Building Contrasting Models and Attributes. *Energy Build.* **2015**, *92*, 322–330.
58. Shi, G.; Liu, D.; Wei, Q. Energy Consumption Prediction of Office Buildings Based on Echo State Networks. *Neurocomputing* **2016**, *216*, 478–488.
59. Jovanović, R.Ž.; Sretenović, A.A.; Živković, B.D. Ensemble of Various Neural Networks for Prediction of Heating Energy Consumption. *Energy Build.* **2015**, *94*, 189–199.
60. Bourhnane, S.; Abid, M.R.; Lghoul, R.; Zine-Dine, K.; Elkamoun, N.; Benhaddou, D. Machine Learning for Energy Consumption Prediction and Scheduling in Smart Buildings. *SN Appl. Sci.* **2020**, *2*, 1–10.
61. Azevedo Guedes, A.L.; Carvalho Alvarenga, J.; Dos Santos Sgarbi Goulart, M.; Rodriguez y Rodriguez, M.V.; Pereira Soares, C.A. Smart cities: The main drivers for increasing the intelligence of cities. *Sustainability* **2018**, *10*, 3121.
62. Aggarwal, T.; Solomon, P. Quantitative Analysis of the Development of Smart Cities in India. *Smart Sustain. Built Environ.* **2019**, *9*, 711–726.
63. Bennon, M.; Kim, M.J.; Levitt, R.E. US Infrastructure Gap(s): Federal Policy and Local Public Institutions. *SSRN* **2017**, 3036650, doi:10.2139/ssrn.3036650.
64. Drozhzhin, S.I.; Shiyan, A.V.; Mityagin, S.A. Smart City Implementation and Aspects: The Case of St. Petersburg. In *Proceedings of the International Conference on Electronic Governance and Open Society: Challenges in Eurasia*; Springer: Berlin/Heidelberg, Germany, 2018; pp. 14–25.
65. Dubey, S.K.; Sharma, D. An Overview of Sustainable Dimensions and Indicators for Smart City. In *Green Technologies and Environmental Sustainabilityl*; Springer: Cham, Switzerland, 2017; pp. 229–240.
66. Geraskina, I.; Kopyrin, A. "Smart City" Concept in the Urban Economy Digitalization System of St. Petersburg. In *Proceedings of the E3S Web of Conferences*; EDP Sciences: Les Ulis, France, 2021; Vol. 281, p. 08006.
67. Lee, J. Evolution of Korean Smart City Programs: Challenges and Opportunities. In Proceedings of the 2020 IEEE Technology & Engineering Management Conference (TEMSCON), Virtual, 3 June 2020; pp. 1–5.
68. Marchetti, D.; Oliveira, R.; Figueira, A.R. Are Global North Smart City Models Capable to Assess Latin American Cities? A Model and Indicators for a New Context. *Cities* **2019**, *92*, 197–207.
69. Mboup, G.; Diongue, M.; Ndiaye, S. Smart city foundation—Driver of smart cities. In *Smart Economy in Smart Cities*; Springer: Berlin/Heidelberg, Germany, 2017; pp. 841–869.
70. Rahman, A.; Maksum, I.R. Initiative of Smart City Development in South Tangerang City: An Approach Toward Sustainable City. In Proceedings of the 2nd ICASPGS International Conference on Administrative Sciences, Policy and Governance Studies, Jakarta, Indonesia, 30–31 October 2018; p. 1.
71. Sotres, P.; Santana, J.R.; Sánchez, L.; Lanza, J.; Muñoz, L. Practical Lessons from the Deployment and Management of a Smart City Internet-of-Things Infrastructure: The Smartsantander Testbed Case. *IEEE Access* **2017**, *5*, 14309–14322.
72. Vershitsky, A.; Egorova, M.; Platonova, S.; Berezniak, I.; Zatsarinnaya, E. Municipal Infrastructure Management Using Smart City Technologies. *Theor. Empir. Res. Urban Manag.* **2021**, *16*, 20–39.
73. Yang, F.; Wen, X.; Aziz, A.; Luhach, A.K. The Need for Local Adaptation of Smart Infrastructure for Sustainable Economic Management. *Environ. Impact Assess. Rev.* **2021**, *88*, 106565.
74. Aslam, S.; Ullah, H.S. A Comprehensive Review of Smart Cities Components, Applications, and Technologies Based on Internet of Things. *arXiv* **2020**, arXiv:2002.01716 2020.
75. Alberti, A.; Senese, M. Developing Capacities for Inclusive and Innovative Urban Governance. In *Governance for Urban Services: Access, Participation, Accountability, and Transparency*; Springer: Singapore, 2020; pp. 127–152.
76. Daraba, D.; Cahaya, A.; Guntur, M.; Aslinda, A.; Akib, H. Strategy of Governance in Transportation Policy Implementation: Case Study of Bus Rapid Transit (BRT) Program in Makassar City. *Acad. Strateg. Manag. J.* **2018**, *17*, 1–12.
77. Izmailova, M.; Veselovskii, M.; Aleksahina, V. Features and Perspectives of Transition of Russian Economy to Technological Paradigm: Regional Aspect. *Adv. Econ. Bus. Manag. Res.* **2017**, *39*, 64–68.
78. Joelsson, I. Risky Urban Futures: The Bridge, the Fund and Insurance in Dar Es Salaam. In *African Cities and Collaborative Futures*; Manchester Universiy Press: Manchester, UK, 2021;pp. 143–165.
79. Krukowska, M. China and Africa. Cooperation Outlook after the 6th FOCAC Summit in Johannesburg, South Africa. *J. Mod. Sci.* **2016**, *3*, 157–180.
80. Leza, S. Innovations in Green Building: Why Sustainable Construction Should Include Localized Knowledge. Ph.D. Thesis, Worcester Polytechnic Institute, Worcester, MA, USA, 2020.
81. Liandi, P.; Yanhui, Z. Study on the Curriculum System of Labor Education in Vocational Colleges under the Background of Dawan District Construction. In Proceedings of the 2019 7th International Education, Economics, Social Science, Arts, Sports and Management Engineering Conference (IEESASM 2019), Dalian, China, 20–21 December 2019.
82. Magare, S.S.; Dudhgaonkar, A.A.; Kondekar, S.R. Security and Privacy Issues in Smart City: Threats and Their Countermeasures. In *Security and Privacy Applications for Smart City Development*; Springer: Berlin/Heidelberg, Germany, 2021; pp. 37–52.
83. Osunsanmi, T.O.; Aigbavboa, C.O.; Oke, A.; Onyia, M.E. Making a Case for Smart Buildings in Preventing Corona-Virus: Focus on Maintenance Management Challenges. *Int. J. Constr. Manag.* **2020**, *20*, 1–10.

84. Plyuschikov, V.G.; Avdotin, V.P.; Arefieva, E.V.; Gurina, R.R.; Bolgov, M.V. Hydrological Risk Management of Urbanized Areas in Framework of the Smart City Concept. *IOP Conf. Ser. Earth Environ. Sci.* **2021**, *691*, 012019.
85. Popoola, O.O.; Alao, S.O.; Akinbowale, K.O. Application of the Smart Cities Approach to Water Management in Ijapo Housing Estate, Akure, Nigeria. In *Inclusive City Growth and the Poor: Policies, Challenges and Prospects*; Community Participation Research Group (COPAREG): Minna, Nigeria, 2018; pp. 38–58.
86. Roman, K. Analysis and Evaluation of the Implementation Level of the Smart City Concept in Selected Polish Cities. *BRAIN Broad Res. Artif. Intell. Neurosci.* **2018**, *9*, 138–145.
87. Sidhu, N.; Pons-Buttazzo, A.; Muñoz, A.; Terroso-Saenz, F. A Collaborative Application for Assisting the Management of Household Plastic Waste through Smart Bins: A Case of Study in the Philippines. *Sensors* **2021**, *21*, 4534.
88. Tripi, K.; Polcari, J.; Hou, R.; Nguyen, T.; Graham, T.; Pottow, C. *National Parks & Technology*; 2020. Available online: https://digital.wpi.edu/downloads/gb19f849v (accessed on 20 September 2021)
89. Vedapradha, R.; Ravi, H.; Rajasekar, A. Blockchain technology: a paradigm shift in investment banking. In *Cryptocurrencies and Blockchain Technology Applications*; Wiley: Hoboken, NJ, USA, 2020; pp. 239–259.
90. Yingqin, Z. China-Nordic Blue Economic Passage: Basis, Challenges and Paths. *China Intl. Stud.* **2019**, *78*, 29.
91. Zhang, F.-S.; Jia, Z.-F. The Smart City Ecological Construction Based on Big Data in FIOT Operation Model. In Proceedings of the 2018 3rd International Conference on Education, Management and Systems Engineering (EMSE 2018), Xiamen, China, 25–26 November 2018.
92. Andrade, R.O.; Yoo, S.G.; Tello-Oquendo, L.; Ortiz-Garcés, I. A Comprehensive Study of the IoT Cybersecurity in Smart Cities. *IEEE Access* **2020**, *8*, 228922–228941.
93. Bellefleur, R.; Wang, D. *IoT-Enabled Smart City Security Considerations and Solutions*; Georgetown University: Washington, DC, USA, 2018.
94. Bulut, U.; Frye, E.; Greene, J.; Li, Y.; Lee, M. The Hidden Price of Convenience: A Cyber-Inclusive Cost-Benefit Analysis of Smart Cities. In *Research in Mathematics and Public Policy*; Springer: Berlin/Heidelberg, Germany, 2020; pp. 81–92.
95. Choo, K.-K.R.; Gai, K.; Chiaraviglio, L.; Yang, Q. A Multidisciplinary Approach to Internet of Things (IoT) Cybersecurity and Risk Management. *Comput. Secur.* **2021**, *102*, 102136.
96. Dissanayake, V.D. A Review of Cyber Security Risks in an Augmented Reality World; University of Sri Lanka, Institute of Information Technology, Malabe, Sri Lanka, 2019.
97. Frick, K.T.; Kumar, T.; Mendonça Abreu, G.K.; Post, A. *Benchmarking "Smart City" Technology Adoption in California: Developing and Piloting a Data Collection Approach*; UC Office of the President: University of California Institute of Transportation Studies: Berkeley, CA, USA, 2021.
98. Gad, M.; Abualhaol, I. Securing Smart Cities Systems and Services: A Risk-Based Analytics-Driven Approach. In *Transportation and Power Grid in Smart Cities*; Communication Networks and Services; Wiley: Hoboken, NJ, USA, 2019; pp. 577–589.
99. Gcaza, N. Cybersecurity Awareness and Education: A Necessary Parameter for Smart Communities. In *Proceedings of the Twelfth International Symposium on Human Aspects of Information Security & Assurance (HAISA)*; University of Plymouth: Plymouth, UK, 2018; pp. 80–90.
100. Gregory, R.R.; Guidera Brown, J. CTIA's IoT Cybersecurity Certification Program May Inform the Future of Transportation and Smart Cities. *J. Robot. Artif. Intell. Law* **2018**, *2*, 177–181.
101. Kalinin, M.; Krundyshev, V.; Zegzhda, P. Cybersecurity Risk Assessment in Smart City Infrastructures. *Machines* **2021**, *9*, 78.
102. Krundyshev, V. Neural Network Approach to Assessing Cybersecurity Risks in Large-Scale Dynamic Networks. In Proceedings of the 13th International Conference on Security of Information and Networks, Merkez, Turkey, 4–7 November 2020; pp. 1–8.
103. Krundyshev, V.; Kalinin, M. The Security Risk Analysis Methodology for Smart Network Environments. In Proceedings of the 2020 International Russian Automation Conference (RusAutoCon), Sochi, Russia, 6–12 September 2020; pp. 437–442.
104. Lakhno, V.; Kasatkin, D.; Blozva, A. Modeling Cyber Security of Information Systems Smart City Based on the Theory of Games and Markov Processes. In Proceedings of the 2019 IEEE International Scientific-Practical Conference Problems of Infocommunications, Science and Technology (PIC S&T), Kyiv, Ukraine, 8–11 October 2019; pp. 497–501.
105. Li, Z.; Liao, Q. Economic Solutions to Improve Cybersecurity of Governments and Smart Cities via Vulnerability Markets. *Gov. Inf. Q.* **2018**, *35*, 151–160.
106. Lim, H.S.M.; Taeihagh, A. Autonomous Vehicles for Smart and Sustainable Cities: An in-Depth Exploration of Privacy and Cybersecurity Implications. *Energies* **2018**, *11*, 1062.
107. Marie-Noëlle Brisson, C.R.E.; Doggendorf, D.; Savoie, M. Cybersecurity of Building Technology: Smart Cities and Smart Buildings Require Smart Protection. *Couns. Real Estate* **2019**, *43*, 1–9.
108. McFarland, R.J.; Olatunbosun, S.B. An Exploratory Study on the Use of Internet_of_medical_things (IoMT) in the Healthcare Industry and Their Associated Cybersecurity Risks. In Proceedings of the International Conference on Internet Computing (ICOMP), The Steering Committee of The World Congress in Computer Science, Computer Engineering, & Applied Computing (CSCE), Las Vegas, NV, USA, 12–15 July 2019; pp. 115–121.
109. Modoni, G.E.; Sacco, M.; Trombetta, A. Strengthening the Cybersecurity of ManufacturingCompanies: A Semantic Approach Compliant with the NISt framework. *ERCIM NEWS* **2018**, *14*, 33–34.
110. Pet Ho, R. Security Issues of Smart City Construction. *Interdiscip. Descr. Complex Syst. INDECS* **2020**, *18*, 337–342.
111. Pet Ho, R.; Tokody, D. Building and Operating a Smart City. *Interdiscip. Descr. Complex Syst. INDECS* **2019**, *17*, 476–484.
112. Pondes, M.A.J. Managing Cyber Risks in Smart Cities. Master's Thesis, University of Twente, Enschede, The Netherlands, 2020.

113. Qamar, T.; Bawany, N.Z. A Cyber Security Ontology for Smart City. *Int. J. Inf. Technol. Secur.* **2020**, *12*, 63–74.
114. Sadik, S.; Ahmed, M.; Sikos, L.F.; Islam, A.K.M. Toward a Sustainable Cybersecurity Ecosystem. *Computers* **2020**, *9*, 74.
115. Serrano, W. The Blockchain Random Neural Network for Cybersecure IoT and 5G Infrastructure in Smart Cities. *J. Netw. Comput. Appl.* **2021**, *175*, 102909.
116. Tang, M.; Yin, J.; Alazab, M.; Cao, J.; Luo, Y. Modeling of Extreme Vulnerability Disclosure in Smart City Industrial Environments. *IEEE Trans. Ind. Inform.* **2020**, *17*, 4150–4158.
117. Teufel, S.; Burri, R.; Teufel, B. Cybersecurity Guideline for the Utility Business a Swiss Approach. In Proceedings of the 2018 International Conference on Smart Grid and Clean Energy Technologies (ICSGCE), Kajang, Malaysia, 29 May–1 June 2018; pp. 1–6.
118. Volkov, S.; Gagliano, A. The Transparency Times: An Installation Promoting Transparency in Urban Sensing. In Proceedings of the 32nd International BCS Human Computer Interaction Conference, Belfast, UK, 4–6 July 2018; pp. 1–5.
119. Vrabie, C. Technological Infrastructure for Building a Smart Ecosystem. In Proceedings of the Strategica International Academic Conference, Bucharest, Romania, 10–11 October 2019; pp. 695–702.
120. Digiesi, S.; Facchini, F.; Mossa, G.; Mummolo, G.; Verriello, R. A Cyber-Based Dss for a Low Carbon Integrated Waste Management System in a Smart City. *IFAC-PapersOnLine* **2015**, *48*, 2356–2361.
121. Aarthi, M.; Bhuvaneshwaran, A. Iot Based Drainage and Waste Management Monitoring and Alert System for Smart City. *Ann. Rom. Soc. Cell Biol.* **2021**, *25*, 6641–6651.
122. Abdullah, N.; Alwesabi, O.A.; Abdullah, R. Iot-Based Smart Waste Management System in a Smart City. In *Proceedings of the International Conference of Reliable Information and Communication Technology*; Springer: Berlin/Heidelberg, Germany, 2018; pp. 364–371.
123. Akbarpour, N.; Salehi-Amiri, A.; Hajiaghaei-Keshteli, M.; Oliva, D. An Innovative Waste Management System in a Smart City under Stochastic Optimization Using Vehicle Routing Problem. *Soft Comput.* **2021**, *25*, 6707–6727.
124. Atayero, A.A.; Popoola, S.I.; Williams, R.; Badejo, J.A.; Misra, S. Smart City Waste Management System Using Internet of Things and Cloud Computing. In *Proceedings of the International Conference on Intelligent Systems Design and Applications*; Springer: Berlin/Heidelberg, Germany, 2019; pp. 601–611.
125. Bakhshi, T.; Ahmed, M. Iot-Enabled Smart City Waste Management Using Machine Learning Analytics. In Proceedings of the 2018 2nd International Conference on Energy Conservation and Efficiency (ICECE), Lahore, Pakistan, 16–17 October 2018; pp. 66–71.
126. Bansal, E.; Goel, A.; Gupta, T.; Piplani, A. Smart City Hygiene Management Using Android Application. *Int. J. Progress. Res. Sci. Eng.* **2020**, *1*, 190–193.
127. Chen, W.-E.; Wang, Y.-H.; Huang, P.-C.; Huang, Y.-Y.; Tsai, M.-Y. A Smart IoT System for Waste Management. In Proceedings of the 2018 1st International Cognitive Cities Conference (IC3), Okinawa, Japan, 7–9 August 2018.
128. Falch, M.; Maestrini, M. Public Private Partnership in Smart City Waste Management-a Business Case. In Proceedings of the 2019 CTTE-FITCE: Smart Cities & Information and Communication Technology (CTTE-FITCE), Ghent, Belgium, 25–27 September 2019; pp. 1–6.
129. Fayomi, G.U.; Mini, S.E.; Chisom, C.M.; Fayomi, O.S.I.; Udoye, N.E.; Agboola, O.; Oomole, D. Smart Waste Management for Smart City: Impact on Industrialization. *IOP Conf. Ser. Earth Environ. Sci.* **2021**, *655*, 012040.
130. Harith, M.Z.M.Z.; Hossain, M.A.; Ahmedy, I.; Idris, M.Y.I.; Soon, T.K.; Noor, R.M. Prototype Development of IoT Based Smart Waste Management System for Smart City. *IOP Conf. Ser. Mater. Sci. Eng.* **2020**, *884*, 012051.
131. Kotzé, P.; Coetzee, L. Opportunities for the Internet of Things in the Water, Sanitation and Hygiene Domain. In Proceedings of the IFIP International Cross-Domain Conference, IFIPIoT 2018 Held at the 24th IFIP World Computer Congress; Poznan, Poland, 18–19 September 2018; pp. 194–210.
132. Mingaleva, Z.; Vukovic, N.; Volkova, I.; Salimova, T. Waste Management in Green and Smart Cities: A Case Study of Russia. *Sustainability* **2020**, *12*, 94.
133. Mokale, P. Smart Waste Management under Smart City Mission—Its Implementation and Ground Realities. *Int. J. Innov. Technol. Explor. Eng.* **2019**, *8*, 3095–3103.
134. Nancy, G.P.; Resmi, R. Smart City Platform Development for Waste Management. *Int. Res. J. Eng. Technol.* **2019**, *6*, 2395–0056.
135. Ng, T.X. Garbage Bin Monitoring for Smart Residence. Ph.D. Thesis, Universiti Tunku Abdul Rahman, Kampar, Malayasia, 2018.
136. Note, F.P. Top 10 Quick-Win Smart City Services for Cairo Governorate: Estimated Budgets and Actions; World Bank Group: Washington, DC, USA, 2018.
137. Onoda, H. Smart Approaches to Waste Management for Post-COVID-19 Smart Cities in Japan. *IET Smart Cities* **2020**, *2*, 89–94.
138. Paulchamy, B.; Alwar, E.B.T.; Anbarasu, K.; Hemalatha, R.; Lavanya, R.; Manasa, K.M. IOT Based Waste Management in Smart City. *Asian J. Appl. Sci. Technol.* **2019**, *2*, 387–394.
139. Thibuy, K.; Thokrairak, S.; Jitngernmadan, P. Holistic Solution Design and Implementation for Smart City Recycle Waste Management Case Study: Saensuk City. In Proceedings of the 2020–5th International Conference on Information Technology (InCIT), Shenyang, China, 19–22 February 2020; pp. 233–237.
140. Wang, S.; Liu, X.; Wang, H.; Hu, Q. A Case Study on Spatio-Temporal Data Mining of Urban Social Management Events Based on Ontology Semantic Analysis. *Sustainability* **2018**, *10*, 2084.

141. Barbosa, M.W.; Vicente, A. de la C.; Ladeira, M.B.; Oliveira, M.P.V. de Managing Supply Chain Resources with Big Data Analytics: A Systematic Review. *Int. J. Logist. Res. Appl.* **2018**, *21*, 177–200.
142. Jindal, F.; Mudgal, S.; Choudhari, V.; Churi, P.P. Emerging Trends in Internet of Things. In Proceedings of the 2018 Fifth HCT Information Technology Trends (ITT), Dubai, UAE, 28–29 November 2018; pp. 50–60.
143. Avdeeva, E.; Davydova, T.; Skripnikova, N.; Kochetova, L. Human Resource Development in the Implementation of the Concept of "Smart Cities". In *Proceedings of the E3S Web of Conferences*; EDP Sciences: Les Ulis, France, 2019; Volume 110, p. 02139.
144. Duan, W.; Nasiri, R.; Karamizadeh, S. Smart City Concepts and Dimensions. In Proceedings of the 2019 7th International Conference on Information Technology: IoT and Smart City, Shanghai, China, 20–23 December 2019; pp. 488–492.
145. Gupta, K.; Hall, R.P. Exploring Smart City Project Implementation Risks in the Cities of Kakinada and Kanpur. *J. Urban Technol.* **2021**, *28*, 155–173.
146. Khanda, S. Challenges Faced by Teacher Counsellors of Secondary Schools in the Smart City Bhubaneswar, Odisha. *Int. J. Res. Soc. Sci.* **2018**, *8*, 327–340.
147. Metsis, S. Cross-Sectoral Data Collaboration Platforms. Master's Thesis, Tallinn University of Technology, Tallinn, Estonia, 2020.
148. Mu, M.; Liu, C. Research on the Construction of O2o E-Commerce and Express Industry Collaborative Development Model in Big Data Environment. In Proceedings of the 2018 International Conference on Internet and e-Business, Taipei, Taiwan, 16–18 May 2018; pp. 29–33.
149. Mukhametov, D. Smart City: From the Metaphor of Urban Development to Innovative City Management. *TEM J.* **2019**, *8*, 1247–1251.
150. Potdar, V.; Chandan, A.; Batool, S.; Patel, N. Big energy data management for smart grids—Issues, challenges and recent developments. In *Smart Cities*; Springer: Berlin/Heidelberg, Germany, 2018; pp. 177–205.
151. Pozdniakova, A.; Velska, I. Setting up the Stage for Smart Sustainable City: International and Ukrainian Context the Role of Smart Solutions. *Acta Innov.* **2020**, *34*, 13–24.
152. Sun, X.; Fan, D.; Li, Q.; Fu, B. Interpretation of the Construction Standard of Smart City Standard Systems. In *AI-Based Services for Smart Cities and Urban Infrastructure*; IGI Global: Hershey, PA, USA, 2021; pp. 91–101.
153. Wang, A. Research on the Development of Jinan Smart Tourism in the Age of Big Data. *J. Phys. Conf. Ser.* **2019**, *1302*, 022011.
154. Xu, L.; Hou, J.; Chen, Z.; Gao, J. Issues and Path Selection of Artificial Intelligence Design Talents Training in Applied Undergraduate Universities in Smart City. In Proceedings of the 20th International Conference on Electronic Business, ICEB'20, Hong Kong, China, 5–8 December 2020; pp. 210–217.
155. Yoo, S.; Kim, Y.; Kim, S. Urban Regeneration Plan for Mid-Sized Cities Deploying the Concept of Smart City-Focused on the US Smart City Challenge in 2015. *J. Archit. Inst. Korea Plan. Des.* **2019**, *35*, 29–40.
156. Zhang, X. Construction for the Smart Old-Age Care in an Age of Longevity: A Literature Review. *IOP Conf. Ser. Earth Environ. Sci.* **2021**, *632*, 052042.
157. Adamuscin, A.; Golej, J.; Panik, M. The Challenge for the Development of Smart City Concept in Bratislava Based on Examples of Smart Cities of Vienna and Amsterdam. *EAI Endorsed Trans. Smart Cities* **2016**, *1*, e5.
158. Alhashmi, S.A.; Al-Sumaiti, A.S.; Hassan, M.W.; Rasheed, M.B.; Rodriguez, S.R.R.; Kumar, R.; Heydarian-Forushani, E. Building Energy Management System: An Overview of Recent Literature Research. In Proceedings of the 2019 Advances in Science and Engineering Technology International Conferences (ASET), Dubai, UAE, 26 March–10 April 2019; pp. 1–5.
159. CEYLAN, R., VELIOĞLU, A., HATIPOĞLU, I., OZBAKIR, A., & ENLIL, Z. BUILDING A SMART COMMUNITY IN KADIKÖY, ISTANBUL. *From efficiency to reduction*, 39. 2019. Available online: https://library.oapen.org/bitstream/handle/20.500.12657/39487/fromefficiency.pdf?sequence=1#page=51 (accessed on 20 September 2021)
160. Chen, H.; Yuan, L.; Jing, G. 5G Boosting Smart Cities Development. In Proceedings of the 2020 2nd International Conference on Artificial Intelligence and Advanced Manufacture (AIAM), Manchester, UK, 15–17 October 2020; pp. 154–157.
161. González-Briones, A.; Hernández, G.; Pinto, T.; Vale, Z.; Corchado, J.M. A Review of the Main Machine Learning Methods for Predicting Residential Energy Consumption. In Proceedings of the 2019 16th International Conference on the European Energy Market (EEM), Ljubljana, Slovenia, 18–20 September 2019; pp. 1–6.
162. KE, M.K.; Kumari, N. A Geo-Spatial Approach for Quantifying Rooftop Photovoltaic Energy Potential in Karnal Smart City, Haryana-A Case Study. *J. Appl. Nat. Sci.* **2021**, *13*, 512–519.
163. Krayem, A.; Al Bitar, A.; Faour, G.; Ahmad, A.; Najem, S. *Beirut as a Smart City: Redefining Urban Energy*; Issam Fares Institute for Public Policy and International Affairs: Beirut, Lebanon, 2019.
164. Kuntsman, A. Rethinking Digital Inevitability: How Digital Futures Might Shape Information Sovereignty, Memory and the Environment. *Сканирование Горизонтов: Роль Информационных Технологий В Будущем Гражданского Общества* **2021**, 1, 72–91.
165. Lewandowska, A.; Chodkowska-Miszczuk, J.; Rogatka, K.; Starczewski, T. Smart Energy in a Smart City: Utopia or Reality? Evidence from Poland. *Energies* **2020**, *13*, 5795.
166. Mohamed, M.A.; Almalaq, A.; Awwad, E.M.; El-Meligy, M.A.; Sharaf, M.; Ali, Z.M. An Effective Energy Management Approach within a Smart Island Considering Water-Energy Hub. *IEEE Trans. Ind. Appl.* **2020**, 1–1, doi:10.1109/TIA.2020.3000704.
167. Monteiro, C.S.; Causone, F.; Cunha, S.; Pina, A.; Erba, S. Addressing the Challenges of Public Housing Retrofits. *Energy Procedia* **2017**, *134*, 442–451.
168. Moura, P.; Moreno, J.I.; López López, G.; Alvarez-Campana, M. IoT Platform for Energy Sustainability in University Campuses. *Sensors* **2021**, *21*, 357.

169. Mutani, G.; Vodano, A.; Pastorelli, M. Photovoltaic Solar Systems for Smart Bus Shelters in the Urban Environment of Turin (Italy). In Proceedings of the 2017 IEEE International Telecommunications Energy Conference (INTELEC), Gold Coast, QL, Australia, 22–26 October 2017; pp. 20–25.
170. Rangaswamy, J.; Kumar, T.; Bhalla, K.; Mishra, V. Building Systems Retrofitted with Building Automation System (BAS): Parametric Design Using TRIZ Methodology and Life Cycle Assessment. In *Green Buildings and Sustainable Engineering*; Springer: Berlin/Heidelberg, Germany, 2019; pp. 203–213.
171. Rendon Restrepo, R.A. Smart Cities and Smart Communities from an Urban-Technological Perspective. Master's Thesis, Politecnico di Milano, Milano, Italy, 2018.
172. Sava, G.N.; Pluteanu, S.; Tanasiev, V.; Patrascu, R.; Necula, H. Integration of BIM Solutions and IoT in Smart Houses. In Proceedings of the 2018 IEEE International Conference on Environment and Electrical Engineering and 2018 IEEE Industrial and Commercial Power Systems Europe (EEEIC/I&CPS Europe), Palermo, Italy, 12–15 June 2018; pp. 1–4.
173. Shahzad, Y.; Javed, H.; Farman, H.; Ahmad, J.; Jan, B.; Zubair, M. Internet of Energy: Opportunities, Applications, Architectures and Challenges in Smart Industries. *Comput. Electr. Eng.* **2020**, *86*, 106739.
174. White, T.; Marchet, F. Digital Social Markets: Exploring the Opportunities and Impacts of Gamification and Reward Mechanisms in Citizen Engagement and Smart City Services. In *How Smart Is Your City?*; Springer: Berlin/Heidelberg, Germany, 2021; pp. 103–125.
175. Yoshida, T.; Yamagata, Y.; Chang, S.; de Gooyert, V.; Seya, H.; Murakami, D.; Jittrapirom, P.; Voulgaris, G. Spatial modeling and design of smart communities. In *Urban Systems Design*; Elsevier: Amsterdam, The Netherlands, 2020; pp. 199–255.
176. Zhang, W.; Hu, W.; Wen, Y. Thermal Comfort Modeling for Smart Buildings: A Fine-Grained Deep Learning Approach. *IEEE Internet Things J.* **2019**, *6*, 2540–2549, doi:10.1109/JIOT.2018.2871461.
177. Aamir, M.; Masroor, S.; Ali, Z.A.; Ting, B.T. Sustainable Framework for Smart Transportation System: A Case Study of Karachi. *Wirel. Pers. Commun.* **2019**, *106*, 27–40.
178. Aldakkhelallah, A.; Simic, M. Autonomous Vehicles in Intelligent Transportation Systems. In *Proceedings of the International Conference on Human-Centered Intelligent Systems*; Springer: Berlin/Heidelberg, Germany, 2021; pp. 185–198.
179. Al-Turjman, F.; Houdjedj, A. Learning in Cities' Cloud-Based IoT. In *The Cloud in IoT-Enabled Spaces*; CRC Press: Boca Raton, FL, USA, 2019; pp. 209–210.
180. Angeline, R.; Ramapuram, C.; Nadu, T.; Eshasree, I.M. Intelligent Transportation System Using IOT. *Int. J. Emerg. Technol. Eng. Res.* **2018**, *6*, 20–25.
181. Anthony Jnr, B. Applying Enterprise Architecture for Digital Transformation of Electro Mobility towards Sustainable Transportation. In Proceedings of the 2020 on Computers and People Research Conference, Nuremberg, Germany, 19–21 June 2020; pp. 38–46.
182. Azgomi, H.F.; Jamshidi, M. A Brief Survey on Smart Community and Smart Transportation. In Proceedings of the 2018 IEEE 30th International Conference on Tools with Artificial Intelligence (ICTAI), Volos, Greece, 5–7 November 2018; pp. 932–939.
183. Dahbour, S.; Qutteneh, R.; Al-Shafie, Y.; Tumar, I.; Hassouneh, Y.; Issa, A.A. Intelligent Transportation System in Smart Cities (ITSSC). In *Proceedings of the SAI Intelligent Systems Conference*; Springer: Berlin/Heidelberg, Germany, 2018; pp. 1157–1170.
184. Dixit, R.S.; Choudhary, S.L. Internet of Things Enabled by Artificial Intelligence. In *Towards Smart World*; Chapman and Hall/CRC: Boca Raton, FL, USA, 2020; pp. 173–196.
185. Fourie, P.J.; Jittrapirom, P.; Binder, R.B.; Tobey, M.B.; Medina, S.O.; Maheshwari, T.; Yamagata, Y. Modeling and design of smart mobility systems. In *Urban Systems Design*; Elsevier: Amsterdam, The Netherlands, 2020; pp. 163–197.
186. Hettikankanama, H.; Vasanthapriyan, S. Integrating Smart Transportation System for a Proposed Smart City: A Mapping Study. In Proceedings of the 2019 International Research Conference on Smart Computing and Systems Engineering (SCSE), Colombo, Sri Lanka, 28 March 2019; pp. 196–203.
187. Jagirdar, R. Development and Evaluation of Low Cost 2-D LiDAR Based Traffic Data Collection Methods. Ph.D. Thesis, New Jersey Institute of Technology, Newark, NJ, USA, 2020.
188. Jan, B.; Farman, H.; Khan, M.; Talha, M.; Din, I.U. Designing a Smart Transportation System: An Internet of Things and Big Data Approach. *IEEE Wirel. Commun.* **2019**, *26*, 73–79.
189. Javed, M.A.; Muram, F.U.; Fattouh, A.; Punnekkat, S. Enforcing Geofences for Managing Automated Transportation Risks in Production Sites. In *European Dependable Computing Conference*; Springer: Berlin/Heidelberg, Germany, 2020; pp. 113–126.
190. Jimenez, J.A. Smart transportation systems. In *Smart Cities*; Springer: Berlin/Heidelberg, Germany, 2018; pp. 123–133.
191. Kelley, S.B.; Lane, B.W.; Stanley, B.W.; Kane, K.; Nielsen, E.; Strachan, S. Smart Transportation for All? A Typology of Recent US Smart Transportation Projects in Midsized Cities. *Ann. Am. Assoc. Geogr.* **2020**, *110*, 547–558.
192. Neilson, A.; Daniel, B.; Tjandra, S. Systematic Review of the Literature on Big Data in the Transportation Domain: Concepts and Applications. *Big Data Res.* **2019**, *17*, 35–44.
193. Nesbitt, K.; Zacherle, B. Elevating Humanity to Optimize the Mobility Revolution. In Proceedings of the International Conference on Sustainable Infrastructure 2019: Leading Resilient Communities through the 21st Century, Los Angeles, CA, USA, 6–9 November 2019; American Society of Civil Engineers: Reston, VA, USA, 2019; pp. 722–728.
194. Nguyen, D.D.; Rohács, J.; Rohács, D.; Boros, A. Intelligent Total Transportation Management System for Future Smart Cities. *Appl. Sci.* **2020**, *10*, 8933.
195. Onwuegbuchunam, D.E.; Ebiringa, O.T.; Etus, C. Framework for Development of Transport Data Repository: A Case of FCT, Abuja, Nigeria. *Educ. Soc. Sci. Humanit. Manag. Sci. J.* **2018**, *18*, 344–349.

196. Perkins, R.; Couto, C.D.; Costin, A. Data Integration and Innovation: The Future of the Construction, Infrastructure, and Transportation Industries. In *Future of Information Exchanges and Interoperability*; Smart Construction Informatics (SCI) Laboratory: Gainesville, FL, USA, 2020.
197. Qoradi, M.D.; Al-Harbi, M.S.; Aina, Y.A. Using GIS-Based Intelligent Transportation Systems in the Enhancement of University Campus Commuting in a Smart City Context. *Arab. J. Geosci.* **2021**, *14*, 1–13.
198. Rashid, A.M.; Yassin, A.A.; Wahed, A.A.A.; Yassin, A.J. Smart City Security: Face-Based Image Retrieval Model Using Gray Level Co-Occurrence Matrix. *J. Inf. Commun. Technol.* **2020**, *19*, 437–458.
199. Ritchie, S.; Rindt, C.; Deeter, D. Technological Innovation and Intelligent Transportation Systems for the US: Perspectives for the 21st Century. In *US Infrastructure*; Routledge: Abingdon, UK, 2019; pp. 37–54.
200. Sharida, A.; Hamdan, A.; Mukhtar, A.-H. Smart cities: The next urban evolution in delivering a better quality of life. In *Toward Social Internet of Things (SIoT): Enabling Technologies, Architectures and Applications*; Springer: Berlin/Heidelberg, Germany, 2020; pp. 287–298.
201. Sperling, J.; Young, S.E.; Garikapati, V.; Duvall, A.L.; Beck, J. *Mobility Data and Models Informing Smart Cities*; National Renewable Energy Lab.: Golden, CO, USA, 2019.
202. Thiranjaya, C.; Rushan, R.; Udayanga, P.; Kaushalya, U.; Rankothge, W. Towards a Smart City: Application of Optimization for a Smart Transportation Management System. In Proceedings of the 2018 IEEE International Conference on Information and Automation for Sustainability (ICIAfS), Colombo, Sri Lanka, 21–22 December 2018; pp. 1–6.
203. Trivedi, P.; Zulkernine, F. Componentry Analysis of Intelligent Transportation Systems in Smart Cities towards a Connected Future. In Proceedings of the 2020 IEEE 22nd International Conference on High Performance Computing and Communications; IEEE 18th International Conference on Smart City; IEEE 6th International Conference on Data Science and Systems (HPCC/SmartCity/DSS), Yanuca Island, Cuvu, Fiji, 14–16 December 2020; pp. 1073–1079.
204. Verma, B.; Snodgrass, R.; Henry, B.; Smith, B.; Daim, T. Smart cities-an analysis of smart transportation management. In *Managing Innovation in a Global and Digital World*; Springer: Berlin/Heidelberg, Germany, 2020; pp. 367–388.
205. Wen, Y.; Zhang, S.; Zhang, J.; Bao, S.; Wu, X.; Yang, D.; Wu, Y. Mapping Dynamic Road Emissions for a Megacity by Using Open-Access Traffic Congestion Index Data. *Appl. Energy* **2020**, *260*, 114357.
206. Yan, J.; Liu, J.; Tseng, F.-M. An Evaluation System Based on the Self-Organizing System Framework of Smart Cities: A Case Study of Smart Transportation Systems in China. *Technol. Forecast. Soc. Chang.* **2020**, *153*, 119371.
207. Bolay, J.-C. When Inclusion Means Smart City: Urban Planning Against Poverty. In *Proceedings of the Future Technologies Conference*; Springer: Berlin/Heidelberg, Germany, 2019; pp. 283–299.
208. Abd El-Mawla, N.; Nagy, A. IoT and Civil Engineering Based Solutions for Global And Environmental Risks of 2019. *Development* **2020**, *20*, 24.
209. Dixon, B.; Johns, R. Vision for a Holistic Smart City-HSC: Integrating Resiliency Framework via Crowdsourced Community Resiliency Information System (CRIS). In Proceedings of the 2nd ACM SIGSPATIAL International Workshop on Advances on Resilient and Intelligent Cities, Chicago, IL, USA, 5 November 2019; pp. 1–4.
210. Garau, C.; Nesi, P.; Paoli, I.; Paolucci, M.; Zamperlin, P. A Big Data Platform for Smart and Sustainable Cities: Environmental Monitoring Case Studies in Europe. In *Proceedings of the International Conference on Computational Science and its Applications*; Springer: Berlin/Heidelberg, Germany, 2020; pp. 393–406.
211. Ghosh, A.; Sarkar, J.P.; Das, B. Sustainable Energy Recovery from Municipal Solid Waste (MSW) Using Bio-Reactor Landfills for Smart City Development. In Proceedings of the 2019 IEEE International Conference on Sustainable Energy Technologies and Systems (ICSETS), Bhubaneswar, India, 26 February–1 March 2019; pp. 242–246.
212. He, Z.; Liu, Z.; Wu, H.; Gu, X.; Zhao, Y.; Yue, X. Research on the Impact of Green Finance and Fintech in Smart City. *Complexity* **2020**, *2020*, 6673386.
213. Ignac-Nowicka, J.; Zarebinska, D.; Kaniak, W. Smart City Idea: Use of People's Opinions on the Environmental Threats for Intelligent Management of the City. *Multidiscip. Asp. Prod. Eng.* **2019**, *2*, 140–150.
214. Liota, C. Sustainable Transport Solutions for the Concept of Smart City: The Case of Umeå. Master's Thesis, Department of Geography and Economic History, Umeå University, Umeå, Sweden, 2018.
215. Mora, L.; Deakin, M.; Zhang, X.; Batty, M.; de Jong, M.; Santi, P.; Appio, F.P. Assembling Sustainable Smart City Transitions: An Interdisciplinary Theoretical Perspective. *J. Urban Technol.* **2020**, *28*, 1–27.
216. Anthony Townsend. Smart Cities: What Do We Need to Know to Plan and Design Them Better?. 25 Oct 2017. Available online: https://items.ssrc.org/parameters/smart-cities-what-do-we-need-to-know-to-plan-and-design-them-better/ (accessed on 20 September 2021)
217. Nica, E.; Konecny, V.; Poliak, M.; Kliestik, T. Big Data Management Of Smart Sustainable Cities: Networked Digital Technologies And Automated Algorithmic Decision-Making Processes. *Manag. Res. Pract.* **2020**, *12*, 48–57.
218. Passarelli, D. The Future of City/Cities. In Economic and Policy Implications of Artificial Intelligence; Springer: Berlin/Heidelberg, Germany, 2020; pp. 161–169.
219. Pfeffer, K.; Verrest, H. Perspectives on the Role of Geo-Technologies for Addressing Contemporary Urban Issues: Implications for IDS. *Eur. J. Dev. Res.* **2016**, *28*, 154–166.
220. Ramirez Lopez, L.J.; Grijalba Castro, A.I. Sustainability and Resilience in Smart City Planning: A Review. *Sustainability* **2021**, *13*, 181.

221. Schmeleva, A.; Bezdelov, S. Environmental Aspects of the Housing Renovation Program in Moscow under Sharing and Circular Economy Conditions. *E3S Web Conf.* **2020**, *203*, 05013.
222. Shohistahon, D.; Rahimboy, D.; Zukhra, K.; Marg'uba, X.; Ganisher, K. Technologies Of "Smart City" in the Republic of Uzbekistan. *Eur. J. Mol. Clin. Med.* **2020**, *7*, 6357–6363.
223. Sovacool, B.K.; Del Rio, D.F.; Griffiths, S. Policy Mixes for More Sustainable Smart Home Technologies. *Environ. Res. Lett.* **2021**, *16*, 054073.
224. Tanda, A.; De Marco, A. Drivers of Public Demand of IoT-Enabled Smart City Services: A Regional Analysis. *J. Urban Technol.* **2018**, *25*, 77–94.
225. Telang, S.; Chel, A.; Nafdey, R.; Kaushik, G. Solar Energy for Sustainable Development of a Smart City. In *Security and Privacy Applications for Smart City Development*; Springer: Berlin/Heidelberg, Germany, 2021; pp. 155–169.
226. Thyagaraj Naidu, S. Energy Harvesting from Human Motion in Smart Cities: An Alternative Renewable Energy Source. Bachelor's Thesis, Tampere University of Applied Sciences, Pirkanmaa, Finland, 2021.
227. Viitanen, J.; Kingston, R. Future City Strategies and Green Growth–Outsourcing Democratic and Environmental Resilience to the Global Technology Sector. *Environ. Plan. A* **2014**, *46*, 803–819.
228. Wang, A.; Lin, W.; Liu, B.; Wang, H.; Xu, H. Does Smart City Construction Improve the Green Utilization Efficiency of Urban Land? *Land* **2021**, *10*, 657.
229. Wang, K.; Yan, F.; Zhang, Y.; Xiao, Y.; Gu, L. Supply Chain Financial Risk Evaluation of Small-and Medium-Sized Enterprises under Smart City. *J. Adv. Transp.* **2020**, *2020*, 8849356.
230. Wei, Y. How Drones and Connected and Autonomous Vehicle Can Reform Smart City Initiatives. In Proceedings of the ISUF 2020 Virtual Conference Proceedings, The 21st Century City, Salt Lake City, UT, USA, 1–4 September 2020; Volume 1.
231. Yigitcanlar, T. Smart City Policies Revisited: Considerations for a Truly Smart and Sustainable Urbanism Practice. *World Technopolis Rev.* **2018**, *7*, 97–112.
232. Zacepins, A.; Kviesis, A.; Komasilovs, V.; Bumanis, N. Model for Economic Comparison of Different Transportation Means in the Smart City. *Balt. J. Mod. Comput.* **2019**, *7*, 354–363.
233. Batchelor, D.; Schnabel, M.A. Smart Heritage in Selected Australian Local Government Smart City Policies. In Proceedings of the Revisiting the Role of Architecture for 'Surviving' Development, 53rd International Conference of the Architectural Science Association (ANZAScA), Roorkee, India, 28–30 November 2019; pp. 28–30.
234. Cazacu, S.; Hansen, N.B.; Schouten, B. Empowerment Approaches in Digital Civics. In Proceedings of the 32nd Australian Conference on Human-Computer Interaction, Sydney, NSW, Australia, 2–4 December 2020; pp. 692–699.
235. De Melo Cartaxo, T.; Castilla, J.M.; Dymet, M.; Hossain, K. Digitalization and Smartening Sustainable City Development: An Investigation from the High North European Cities. *Smart Cities Reg. Dev. J.* **2021**, *5*, 83–101.
236. Dueñas, K. *Smart Cities and The Third Industrial Revolution*; University of Nevada: Las Vegas, CA, USA, 2019.
237. Hall, R.E.; Bowerman, B.; Braverman, J.; Taylor, J.; Todosow, H.; Von Wimmersperg, U. *The Vision of a Smart City*; Brookhaven National Lab.: Upton, NY, USA, 2000.
238. Kelkar, N.P.; Spinelli, G. Building Social Capital through Creative Placemaking. *Strateg. Des. Res. J.* **2016**, *9*, 54–66.
239. Klusacek, P.; Konecny, O.; Zgodova, A.; Navratil, J. Application of the Smart City Concept in Process of Urban Recycling-Case Study of Špitálka in Brno, Czech Republic. Deturope Cent. *Eur. J. Reg. Dev. Tour* **2020**, *12*, 22–40.
240. Maalsen, S.; Burgoyne, S.; Tomitsch, M. Smart-Innovative Cities and the Innovation Economy: A Qualitative Analysis of Local Approaches to Delivering Smart Urbanism in Australia. *J. Des. Bus. Soc.* **2018**, *4*, 63–82.
241. Musa, S. Smart Cities-a Road Map for Development. *IEEE Potentials* **2018**, *37*, 19–23.
242. Romanelli, M. Towards Cities as Communities. In *ICT for an Inclusive World*; Springer: Berlin/Heidelberg, Germany, 2020; pp. 125–132.
243. Suciu, G.; Tudor, A.-M. Promoting A Business Through Events In A Smart City. *ELearning Softw. Educ.* **2020**, *3*, 150–155.
244. Tyas, W.P.; Nugroho, P.; Sariffuddin, S.; Purba, N.G.; Riswandha, Y.; Sitorus, G.H.I. Applying Smart Economy of Smart Cities in Developing World: Learnt from Indonesia's Home Based Enterprises. *IOP Conf. Ser. Earth Environ. Sci.* **2019**, *248*, 012078.
245. Jararweh, Y.; Otoum, S.; Al Ridhawi, I. Trustworthy and Sustainable Smart City Services at the Edge. *Sustain. Cities Soc.* **2020**, *62*, 102394.
246. Allam, Z.; Jones, D.S. On the Coronavirus (COVID-19) Outbreak and the Smart City Network: Universal Data Sharing Standards Coupled with Artificial Intelligence (AI) to Benefit Urban Health Monitoring and Management. *Healthcare* **2020**, *8*, 46, doi:10.3390/healthcare8010046.
247. Nuseir, M.T.; Basheer, M.F.; Aljumah, A. Antecedents of Entrepreneurial Intentions in Smart City of Neom Saudi Arabia: Does the Entrepreneurial Education on Artificial Intelligence Matter? *Cogent Bus. Manag.* **2020**, *7*, 1825041.
248. Keymolen, E.; Voorwinden, A. Can We Negotiate? Trust and the Rule of Law in the Smart City Paradigm. *Int. Rev. Law Comput. Technol.* **2020**, *34*, 233–253.
249. Falco, G. Participatory AI: Reducing AI Bias and Developing Socially Responsible AI in Smart Cities. In Proceedings of the 2019 IEEE International Conference on Computational Science and Engineering (CSE) and IEEE International Conference on Embedded and Ubiquitous Computing (EUC), New York, NY, USA, 1–3 August 2019; pp. 154–158.
250. Hwang, J.-S. The evolution of smart city in South Korea: The smart city winter and the city-as-a-platform. In *Smart Cities in Asia*; Edward Elgar Publishing: Cheltenham, UK, 2020.

251. Ahmad, K.; Maabreh, M.; Ghaly, M.; Khan, K.; Qadir, J.; Al-Fuqaha, A. Developing Future Human-Centered Smart Cities: Critical Analysis of Smart City Security, Interpretability, and Ethical Challenges. *arXiv* **2020**, arXiv:2012.09110.
252. Al Ridhawi, I.; Otoum, S.; Aloqaily, M.; Boukerche, A. Generalizing AI: Challenges and Opportunities for Plug and Play AI Solutions. *IEEE Netw.* **2020**, *35*, 372–379.
253. Cugurullo, F. Urban Artificial Intelligence: From Automation to Autonomy in the Smart City. *Front. Sustain. Cities* **2020**, *2*, 38.
254. Fink, J. Digital City Testbed Center: Using Campuses as Smart City Testbeds in the Binational Cascadia Region. In Proceedings of the 2020 IEEE International Conference on Smart Computing (SMARTCOMP), Bologna, Italy, 14–17 September 2020; pp. 362–367.
255. Habib, A.; Alsmadi, D.; Prybutok, V.R. Factors That Determine Residents' Acceptance of Smart City Technologies. *Behav. Inf. Technol.* **2020**, *39*, 610–623.
256. Hassan, R.J.; Zeebaree, S.R.; Ameen, S.Y.; Kak, S.F.; Sadeeq, M.A.; Ageed, Z.S.; Adel, A.-Z.; Salih, A.A. State of Art Survey for Iot Effects on Smart City Technology: Challenges, Opportunities, and Solutions. *Asian J. Res. Comput. Sci.* **2021**, *8*, 32–48.
257. Kuberkar, S.; Singhal, T.K. Factors Influencing Adoption Intention of AI Powered Chatbot for Public Transport Services within a Smart City. *Int. J. Emerg. Technol. Learn.* **2020**, *11*, 948–958.
258. Law, J.W. Application of Artificial Intelligence In Smart City. Ph.D. Thesis, Universiti Tunku Abdul Rahman, Kampar, Malaysia, 2021.
259. Nikitas, A.; Michalakopoulou, K.; Njoya, E.T.; Karampatzakis, D. Artificial Intelligence, Transport and the Smart City: Definitions and Dimensions of a New Mobility Era. *Sustainability* **2020**, *12*, 2789.
260. Olszewski, R.; Palka, P.; Turek, A.; Kietlińska, B.; Platkowski, T.; Borkowski, M. Spatiotemporal Modeling of the Smart City Residents' Activity with Multi-Agent Systems. *Appl. Sci.* **2019**, *9*, 2059.
261. Pereira, G.V.; Wimmer, M.; Ronzhyn, A. Research Needs for Disruptive Technologies in Smart Cities. In Proceedings of the 13th International Conference on Theory and Practice of Electronic Governance, Athens, Greece, 1–3 April 2020; pp. 620–627.
262. Qolomany, B.; Mohammed, I.; Al-Fuqaha, A.; Guizani, M.; Qadir, J. Trust-Based Cloud Machine Learning Model Selection for Industrial IoT and Smart City Services. *IEEE Internet Things J.* **2020**, *8*, 2943–2958.
263. Rahman, M.A.; Hossain, M.S.; Showail, A.J.; Alrajeh, N.A.; Alhamid, M.F. A Secure, Private, and Explainable IoHT Framework to Support Sustainable Health Monitoring in a Smart City. *Sustain. Cities Soc.* **2021**, *72*, 103083.
264. Sakellarides, C. From Viral City to Smart City: Learning from Pandemic Experiences. *Acta Med. Port.* **2020**, *33*, 359–361.
265. Zhong, Y.; Sun, L.; Ge, C. Key Technologies and Development Status of Smart City. *J. Phys. Conf. Ser.* **2021**, *1754*, 012102.
266. Mora-Sánchez, O.B.; López-Neri, E.; Cedillo-Elias, E.J.; Aceves-Martínez, E.; Larios, V.M. Validation of IoT Infrastructure for the Construction of Smart Cities Solutions on Living Lab Platform. *IEEE Trans. Eng. Manag.* **2020**, *68*, 899–908.
267. Nassar, M.A.; Luxford, L.; Cole, P.; Oatley, G.; Koutsakis, P. The Current and Future Role of Smart Street Furniture in Smart Cities. *IEEE Commun. Mag.* **2019**, *57*, 68–73.
268. Mhlanga, D. Artificial Intelligence in the Industry 4.0, and Its Impact on Poverty, Innovation, Infrastructure Development, and the Sustainable Development Goals: Lessons from Emerging Economies? *Sustainability* **2021**, *13*, 5788.
269. Park, E.; Del Pobil, A.P.; Kwon, S.J. The Role of Internet of Things (IoT) in Smart Cities: Technology Roadmap-Oriented Approaches. *Sustainability* **2018**, *10*, 1388, doi:10.3390/su10051388.
270. Apró, D.; Tóth, R.; Orova, M.; Kiss, I. Can Smart City Tools Support Historical Cities Become More Resilient and Regenerative? In Proceedings of the 32TH International Conference on Passive and Low Energy Architecture, Los Angeles, NV, USA, 11–13 July 2016.
271. Abdalla, W.; Renukappa, S.; Suresh, S.; Al-Janabi, R. Challenges for Managing Smart Cities Initiatives: An Empirical Study. In Proceedings of the 2019 3rd International Conference on Smart Grid and Smart Cities (ICSGSC), Berkeley, CA, USA, 25–28 June 2019; pp. 10–17.
272. Okai, E.; Feng, X.; Sant, P. Smart Cities Survey. In Proceedings of the 2018 IEEE 20th International Conference on High Performance Computing and Communications, IEEE 16th International Conference on Smart City; IEEE 4th International Conference on Data Science and Systems (HPCC/SmartCity/DSS), Exeter, UK, 28–30 June 2018; pp. 1726–1730.
273. Wang, P.; Ali, A.; Kelly, W. Data Security and Threat Modeling for Smart City Infrastructure. In Proceedings of the 2015 International Conference on Cyber Security of Smart Cities, Industrial Control System and Communications (SSIC), Shanghai, China, 5–7 August 2015; pp. 1–6.
274. Kitchin, R. *Getting Smarter about Smart Cities: Improving Data Privacy and Data Security*; Data Protection Unit, Department of the Taoiseach, Dublin, Ireland, 2016.
275. Horwitz, J. China Passes New Personal Data Privacy Law, to TAKE effect Nov. 1. Reuters. 2021. Available online: https://www.reuters.com/world/china/china-passes-new-personal-data-privacy-law-take-effect-nov-1-2021-08-20/ (accessed on 4 September 2021).
276. Khan, Z.; Pervez, Z.; Ghafoor, A. Towards Cloud Based Smart Cities Data Security and Privacy Management. In Proceedings of the 2014 IEEE/ACM 7th International Conference on Utility and Cloud Computing, London, UK, 8–11 December 2014; pp. 806–811.
277. Pasqualetti, F.; Dörfler, F.; Bullo, F. Attack detection and identification in cyber-physical systems. *IEEE Trans. Autom. Control.* 2013, 58, 2715–2729.
278. Carl, G.; Kesidis, G.; Brooks, R.R.; Rai, S. Denial-of-service attack-detection techniques. *IEEE Internet Comput.* **2006**, *10*, 82–89.

279. Salem, M.; Hershkop, S.; Stolfo, S. A Survey of Insider Attack Detection Research. *Insider Attack Cyber Secur.* **2008**, *39*, 69–90, doi:10.1007/978-0-387-77322-3_5.
280. Liagkou, V.; Kavvadas, V.; Chronopoulos, S.; Tafiadis, D.; Christofilakis, V.; Peppas, K. Attack Detection for Healthcare Monitoring Systems Using Mechanical Learning in Virtual Private Networks over Optical Transport Layer Architecture. *Computation* **2019**, *7*, 24, doi:10.3390/computation7020024.
281. Nirde, K.; Mulay, P.S.; Chaskar, U.M. IoT Based Solid Waste Management System for Smart City. In Proceedings of the 2017 International Conference on Intelligent Computing and Control Systems (ICICCS), Madurai, India, 15–16 June 2017; pp. 666–669.
282. Malapur, B.S.; Pattanshetti, V.R. IoT Based Waste Management: An Application to Smart City. In Proceedings of the 2017 International Conference on Energy, Communication, Data Analytics and Soft Computing (ICECDS), Chennai, India, 1–2 August 2017; pp. 2476–2486.
283. Sharmin, S.; Al-Amin, S.T. A Cloud-Based Dynamic Waste Management System for Smart Cities. In Proceedings of the 7th Annual Symposium on Computing for Development, Nairobi, Kenya, 18–20 November 2016; pp. 1–4.
284. Szmuc, T.; Kotulski, L.; Wojszczyk, B.; Sedziwy, A.; Donnellan, B.; Lopes, J.A.P.; Martins, J.; Filipe, J. Green AGH Campus. In Proceedings of the 1st International Conference on Smart Grids and Green IT Systems (SMARTGREENS-2012), Porto, Portugal, 19–20 April 2012; pp. 159–162.
285. Nick, G.A.; Pongracz, F. Hungarian Smart Cities Strategies Towards Industry 4.0. *Industry 4.0* **2016**, *1*, 122–127.
286. Novák, J. Pruuzkum Trhu pro Aplikace Internetu Věcí. Bachelor of Science Thesis, České Vysoké Učení Technické v Praze. Vypočetní a Informační Centrum, Prague, Czech Republic, 2018.
287. Jiang, J. What is a Smart City? What are the Difficulties in the Construction Of Smart Cities? Available online: https://zhuanlan.zhihu.com/p/353917160 (accessed on 22 July 2021).
288. ACM. Glossary Smart Devices What Are Smart Devices?. ACM. 2021. Available online: https://www.arm.com/glossary/smart-devices (accessed on 4 September 2021).
289. Solanki, V.K.; Katiyar, S.; BhaskarSemwal, V.; Dewan, P.; Venkatasen, M.; Dey, N. Advanced Automated Module for Smart and Secure City. *Procedia Comput. Sci.* **2016**, *78*, 367–374.
290. Khansari, N.; Mostashari, A.; Mansouri, M. Conceptual Modeling of the Impact of Smart Cities on Household Energy Consumption. *Procedia Comput. Sci.* **2014**, *28*, 81–86.
291. Jettanasen, C.; Songsukthawan, P.; Ngaopitakkul, A. Development of Micro-Mobility Based on Piezoelectric Energy Harvesting for Smart City Applications. *Sustainability* **2020**, *12*, 2933.
292. Şerban, A.C.; Lytras, M.D. Artificial Intelligence for Smart Renewable Energy Sector in Europe—Smart Energy Infrastructures for Next Generation Smart Cities. *IEEE Access* **2020**, *8*, 77364–77377, doi:10.1109/ACCESS.2020.2990123.
293. Crowe, C. Mississippi Power plans 'smart neighborhood' with Tesla Solar Roofs. Smart Cities Dive. 2021. Available online: https://www.smartcitiesdive.com/news/mississippi-power-plans-smart-neighborhood-with-tesla-solar-roofs/589837/ (accessed on 4 September 2021).
294. Anbari, S.; Majidi, B.; Movaghar, A. 3D Modeling of Urban Environment for Efficient Renewable Energy Production in the Smart City. In Proceedings of the 2019 7th Iranian Joint Congress on Fuzzy and Intelligent Systems (CFIS), Bojnord, Iran, 29–31 January 2019; pp. 1–4.
295. Northfield, R. Greening the Smart City. *Eng. Technol.* **2016**, *11*, 38–41.
296. Satyakrishna, J.; Sagar, R.K. Analysis of Smart City Transportation Using IoT. In Proceedings of the 2018 2nd International Conference on Inventive Systems and Control (ICISC), Coimbatore, India, 19–20 January 2018; pp. 268–273.
297. Saarika, P.S.; Sandhya, K.; Sudha, T. Smart Transportation System Using IoT. In Proceedings of the 2017 International Conference On Smart Technologies For Smart Nation (SmartTechCon), Bengaluru, India, 17–19 August 2017; pp. 1104–1107.
298. Menouar, H.; Guvenc, I.; Akkaya, K.; Uluagac, A.S.; Kadri, A.; Tuncer, A. UAV-Enabled Intelligent Transportation Systems for the Smart City: Applications and Challenges. *IEEE Commun. Mag.* **2017**, *55*, 22–28.
299. Lingli, J. Smart City, Smart Transportation: Recommendations of the Logistics Platform Construction. In Proceedings of the 2015 International Conference on Intelligent Transportation, Big Data and Smart City, Halong Bay, Vietnam, 19–20 December 2015; pp. 729–732.
300. Saroj, A.; Roy, S.; Guin, A.; Hunter, M.; Fujimoto, R. Smart City Real-Time Data-Driven Transportation Simulation. In Proceedings of the 2018 Winter Simulation Conference (WSC), Gothenburg, Sweden, 9–12 December 2018; pp. 857–868.
301. Prokshits, E.E.; Zolotukhina, I.A. Energy Efficient Way of Organizing the Urban Environment to Achieve the Goals of Sustainable Development of Areas. In Proceedings of the Поколение будущего: Взгляд молодых ученых-2018, Kursk, Russia, 13–14 November 2018; pp. 117–120.
302. Lugo Santiago, J.A. From City to Smart City: Key Drivers of Change. In *Leadership and Strategic Foresight in Smart Cities*; Springer: Berlin/Heidelberg, Germany, 2020; pp. 21–32.
303. Creutzig, F.; Franzen, M.; Moeckel, R.; Heinrichs, D.; Nagel, K.; Nieland, S.; Weisz, H. Leveraging Digitalization for Sustainability in Urban Transport. *Glob. Sustain.* **2019**, *2*, E14.
304. Kitchin, R. The Real-Time City? Big Data and Smart Urbanism. *GeoJournal* **2014**, *79*, 1–14.
305. Crittenden, C. "A Drama in Time": How Data and Digital Tools Are Transforming Cities and Their Communities. *City Community* **2017**, *16*, 3–8.

306. Jiao, W. The Role of Big Data in Smart City Planning. Ph.D. Thesis, Auckland University of Technology, Auckland, New Zealand, 2018.
307. Gascó-Hernandez, Building a Smart City: Lessons from Barcelona, Information Policy. 2018. Available online: https://www.i-policy.org/2018/05/building-a-smart-city-lessons-from-barcelona.html (accessed on 20 September 2021)
308. Jabbar, M.A.; Prasad, K.; Aluvalu, R. Reimagining the Indian Healthcare Ecosystem with AI for a Healthy Smart City. In *Emerging Technologies in Data Mining and Information Security*; Springer: Singapore, 2021; pp. 543–551.
309. Hölbl, M.; Kompara, M.; Kamišalić, A.; Nemec Zlatolas, L. A systematic review of the use of blockchain in healthcare. *Symmetry* **2018**, *10*, 470.
310. Kong, L. A Study on the AI-Based Online Triage Model for Hospitals in Sustainable Smart City. *Future Gener. Comput. Syst.* **2021**, *125*, 59–70.
311. Gandhi, S.L. Smart Education Service Model Based on IOT Technology. In Proceedings of the International Interdisciplinary Conference on Science Technology Engineering Management Pharmacy and Humanities, Singapore, 22–23 April 2017.
312. Mahmood, S.; Palaniappan, S.; Hasan, R.; Sarker, K.U.; Abass, A.; Rajegowda, P.M. Raspberry PI and Role of IoT in Education. In Proceedings of the 2019 4th MEC International Conference on Big Data And Smart City (ICBDSC), Muscat, Oman, 15–16 January 2019; pp. 1–6.
313. Thelisson, E.; Padh, K.; Celis, L.E. Regulatory mechanisms and algorithms towards trust in AI/ML. In Proceedings of the IJCAI 2017 Workshop on Explainable Artificial Intelligence (XAI), Melbourne, Australia, 19–21 August 2017.
314. Schmidt, P.; Biessmann, F.; Teubner, T. Transparency and trust in artificial intelligence systems. *J. Decis. Syst.* **2020**, *29*, 260–278.
315. Stone, P.; Brooks, R.; Brynjolfsson, E.; Calo, R.; Etzioni, O.; Hager, G.; Hirschberg, J.; Kalyanakrishnan, S.; Kamar, E.; Kraus, S.; et al. Artificial intelligence and life in 2030: The one hundred year study on artificial intelligence; Stanford University: Stanford, CA, USA, 2016.
316. Poola, I. How artificial intelligence in impacting real life everyday. *Int. J. Adv. Res. Dev.* **2017**, *2*, 96–100.
317. Kamble, R.; Deepali, S. Applications of artificial intelligence in human life. *Int. J. Res.-Granthaalayah* **2018**, *6*, 178–188.
318. Ady, L.; Hudasi, L.F. Artificial Intelligence Usage Opportunities in Smart City Data Management. *Interdiscip. Descr. Complex Syst. INDECS* **2020**, *18*, 382–388.
319. Sebastian, A.; Sivagurunathan, S.; Ganeshan, V.M. IoT challenges in data and citizen-centric smart city governance. In *Smart Cities*; Springer: Berlin/Heidelberg, Germany, 2018; pp. 127–151.
320. The Second Half of The AEC Industry. 2020. Available online: https://www.solearth.com/news/news/2020/06/the-second-half-of-the-aec-industry/ (accessed on 20 September 2021).
321. Kijewska, K.; Torbacki, W.; Iwan, S. Application of AHP and DEMATEL Methods in Choosing and Analysing the Measures for the Distribution of Goods in Szczecin Region. *Sustainability* **2018**, *10*, 2365.
322. Tarei, P.K.; Thakkar, J.J.; Nag, B. A Hybrid Approach for Quantifying Supply Chain Risk and Prioritizing the Risk Drivers: A Case of Indian Petroleum Supply Chain. *J. Manuf. Technol. Manag.* **2018**, *29*, 533–569.
323. Radziejowska, A.; Sobotka, B. Analysis of the Social Aspect of Smart Cities Development for the Example of Smart Sustainable Buildings. *Energies* **2021**, *14*, 4330.
324. Kao, Y.-S.; Nawata, K.; Huang, C.-Y. An Exploration and Confirmation of the Factors Influencing Adoption of IoT-Based Wearable Fitness Trackers. *Int. J. Environ. Res. Public Health* **2019**, *16*, 3227.
325. Song, W.; Zhu, Y.; Zhao, Q. Analyzing Barriers for Adopting Sustainable Online Consumption: A Rough Hierarchical DEMATEL Method. *Comput. Ind. Eng.* **2020**, *140*, 106279.
326. Tseng, M.-L.; Lin, C.-W.R.; Sujanto, R.Y.; Lim, M.K.; Bui, T.-D. Assessing Sustainable Consumption in Packaged Food in Indonesia: Corporate Communication Drives Consumer Perception and Behavior. *Sustainability* **2021**, *13*, 8021.
327. Ramesh, K.T.; Sarmah, S.P.; Tarei, P.K. An Integrated Framework for the Assessment of Inbound Supply Risk and Prioritization of the Risk Drivers: A Real-Life Case on Electronics Supply Chain. *Benchmarking Int. J.* **2019**, *27*, 1261–1286.
328. Chen, W.K.; Nalluri, V.; Hung, H.C.; Chang, M.C.; Lin, C.T. Apply DEMATEL to Analyzing Key Barriers to Implementing the Circular Economy: An Application for the Textile Sector. *Appl. Sci.* **2021**, *11*, 3335.
329. Raghuvanshi, J.; Agrawal, R.; Ghosh, P.K. Analysis of barriers to women entrepreneurship: The DEMATEL approach. *J. Entrep.* **2017**, *26*, 220–238.
330. Si, S.L.; You, X.Y.; Liu, H.C.; Zhang, P. DEMATEL technique: A systematic review of the state-of-the-art literature on methodologies and applications. *Math. Probl. Eng.* **2018**, *2018*, 3696457.
331. Ngan, S.L.; How, B.S.; Teng, S.Y.; Leong, W.D.; Loy, A.C.M.; Yatim, P.; Lam, H.L. A hybrid approach to prioritize risk mitigation strategies for biomass polygeneration systems. *Renew. Sustain. Energy Rev.* **2020**, *121*, 109679.
332. Razmjoo, A.; Østergaard, P.A.; Denai, M.; Nezhad, M.M.; Mirjalili, S. Effective Policies to Overcome Barriers in the Development of Smart Cities. *Energy Res. Soc. Sci.* **2021**, *79*, 102175.
333. Mohanty, S.P.; Choppali, U.; Kougianos, E. Everything You Wanted to Know about Smart Cities: The Internet of Things Is the Backbone. *IEEE Consum. Electron. Mag.* **2016**, *5*, 60–70.
334. Perlroth, N.; Sanger, D.E. Cyberattacks Put Russian Fingers on the Switch at Power Plants, US Says. *New York Times, 15 Oct* **2018**. Available online: https://www.nytimes.com/2018/03/15/us/politics/russia-cyberattacks.html (accessed on 20 September 2021)
335. Kosowatz, J. 10 Smart Cities. *Mech. Eng.* **2020**, *142*, 32–37.